\documentclass[12pt]{article}
\usepackage{graphicx}
\usepackage{amsmath,amsthm,amssymb}
\usepackage{color}

\def\empile#1\above#2{\mathrel{\mathop{\kern 0pt#1}\limits_{#2}}}

\newcommand{\sll}{\raise.15ex\hbox{$/$}\kern-.43em\hbox{$l$}}
\newcommand{\slepsilon}{\raise.15ex\hbox{$/$}\kern-.53em\hbox{$\epsilon$}}
\newcommand{\slvarepsilon}{\raise.15ex\hbox{$/$}\kern-.53em\hbox{$\varepsilon$}}
\newcommand{\slL}{\raise.15ex\hbox{$/$}\kern-.53em\hbox{$L$}}
\newcommand{\slP}{\raise.15ex\hbox{$/$}\kern-.53em\hbox{$P$}}
\newcommand{\slp}{\raise.1ex\hbox{$/$}\kern-.63em\hbox{$p$}}
\newcommand{\slq}{\raise.1ex\hbox{$/$}\kern-.53em\hbox{$q$}}
\newcommand{\slv}{\raise.1ex\hbox{$/$}\kern-.63em\hbox{$v$}}
\newcommand{\slR}{\raise.15ex\hbox{$/$}\kern-.53em\hbox{$R$}}
\newcommand{\slQ}{\raise.15ex\hbox{$/$}\kern-.53em\hbox{$Q$}}
\newcommand{\slK}{\raise.15ex\hbox{$/$}\kern-.53em\hbox{$K$}}
\newcommand{\slk}{\raise.15ex\hbox{$/$}\kern-.53em\hbox{$k$}}
\newcommand{\slSigma}{\raise.15ex\hbox{$/$}\kern-.53em\hbox{$\Sigma$}}
\newcommand{\slcalP}{\raise.15ex\hbox{$/$}\kern-.63em\hbox{$\cal P$}}
\newcommand{\slA}{\raise.15ex\hbox{$/$}\kern-.73em\hbox{$A$}}
\newcommand{\slbfA}{\raise.15ex\hbox{$/$}\kern-.73em\hbox{${\imb A}$}}
\newcommand{\slpartial}{\raise.15ex\hbox{$/$}\kern-.53em\hbox{$\partial$}}
\newcommand{\sla}{\raise.15ex\hbox{$/$}\kern-.53em\hbox{$a$}}
\newcommand{\slb}{\raise.15ex\hbox{$/$}\kern-.53em\hbox{$b$}}
\newcommand{\slc}{\raise.15ex\hbox{$/$}\kern-.53em\hbox{$c$}}
\newcommand{\slD}{\raise.15ex\hbox{$/$}\kern-.53em\hbox{$D$}}
\newcommand{\slC}{\raise.15ex\hbox{$/$}\kern-.53em\hbox{$C$}}

\def\bs{\boldsymbol}

\catcode`\@=11

\newcount\@tempcntc
\def\@citex[#1]#2{\if@filesw\immediate\write\@auxout{\string\citation{#2}}\fi
  \@tempcnta\z@\@tempcntb\m@ne\def\@citea{}\@cite{%
        \@for\@citeb:=#2\do%
    {\@ifundefined{b@\@citeb}%
        {\@citeo\@tempcntb\m@ne\@citea%
                \def\@citea{,\penalty\@m\ }{\bf ?}\@warning%
                {Citation `\@citeb' on page \thepage \space undefined}}%
        {\setbox\z@\hbox{\global\@tempcntc0\csname b@\@citeb\endcsname\relax}
     \ifnum\@tempcntc=\z@ \@citeo\@tempcntb\m@ne%
       \@citea\def\@citea{,\penalty\@m}%
       \hbox{\csname b@\@citeb\endcsname}%
     \else%
      \advance\@tempcntb\@ne%
      \ifnum\@tempcntb=\@tempcntc%
      \else\advance\@tempcntb\m@ne\@citeo%
      \@tempcnta\@tempcntc\@tempcntb\@tempcntc\fi\fi}}\@citeo}{#1}}

\def\@citeo{\ifnum\@tempcnta>\@tempcntb\else\@citea
  \def\@citea{,\penalty\@m}%
  \ifnum\@tempcnta=\@tempcntb\the\@tempcnta\else
   {\advance\@tempcnta\@ne\ifnum\@tempcnta=\@tempcntb \else
\def\@citea{--}\fi
    \advance\@tempcnta\m@ne\the\@tempcnta\@citea\the\@tempcntb}\fi\fi}

\catcode`\@=12

\begin{document}

\thispagestyle{empty}
\title {\bf Leptons from heavy-quark semileptonic decay in pA collisions
within the CGC framework}

\author{Hirotsugu Fujii$^1$ and Kazuhiro Watanabe$^2$\footnote{Present address : {Physics Department, Old Dominion University, Norfolk, VA 23529, United States}~\&~{Theory Center, Jefferson Lab, Newport News, VA 23606, United States}}}
\maketitle
\begin{center}
$^1${Institute of Physics, University of Tokyo, Komaba, Tokyo 153-8902, Japan}
\end{center}
\begin{center}
$^2${Key Laboratory of Quark and Lepton Physics (MOE) and Institute of Particle Physics, Central China Normal University, Wuhan 430079, China}
\end{center}

\begin{abstract}
We study single lepton production from semileptonic decays of heavy flavor hadrons ($D,B\rightarrow~l$) in pp and p$A$ collisions at RHIC and the LHC within the saturation/Color-Glass-Condensate (CGC) framework. Using the gluon distribution function obtained with the dipole amplitude, whose energy dependence is described by the Balitsky-Kovchegov equation with running coupling effect, we compute the transverse-momentum ($p_\perp$) spectra of the lepton yields at mid and forward rapidities. We find that a large fraction of leptons at low $p_\perp$ stems from the saturation regime of the incoming gluons in the target, especially in p$A$ collisions at the LHC. The resultant $p_\perp$ spectra is slightly harder than the data, but the nuclear modification factor seems consistent with the data within some uncertainty. We also update the nuclear modification factors for J/$\psi$ and $D$ meson at the LHC energy.
\end{abstract}

\section{Introduction}
Heavy quark production in hadronic collisions has received much attention in studying aspects of quantum chromodynamics (QCD) because its large mass sets the scale for perturbative analyses~\cite{Andronic:2015wma}. 
In recent years, productions of quarkonia and open heavy flavor mesons in proton-nucleus (p$A$) collisions are investigated from the viewpoint of the gluon saturation in hadrons~\cite{Gribov:1984tu,Mueller:1985wy} 
by taking into account the nonlinear evolution effects in the Color-Glass-Condensate (CGC) formalism~\cite{Fujii:2006ab,Fujii:2007ty,Fujii:2013gxa,Fujii:2013yja,Ma:2015sia,Ducloue:2015gfa}
(see also~\cite{Kharzeev:2003sk,Kharzeev:2005zr,Kovchegov:2006qn,Kharzeev:2008nw}). 
High-energy p$A$ collisions can be regarded as a good laboratory for gluon saturation physics:
First, the saturation momentum $Q_s$, which separates the nonlinear regime ($Q<Q_s$) from the linear regime ($Q_s \ll Q$) of a QCD process with an external scale $Q$, is enhanced in the heavy nucleus and at high energies. This is because the saturation scale is expected to be proportional basically to the gluon density per unit transverse area of the target nucleus and scales as $Q_s^2 \propto A^{1/3} x^{-\lambda}$ with Bjorken's $x$ being a longitudinal momentum fraction carried by a gluon and with an empirical parameter $\lambda\sim 0.3$~\cite{GolecBiernat:1998js}.  
Second, it is conventionally assumed that no hot and dense quark-gluon medium is formed in p$A$ collisions, and thus one can study the initial state effects more directly there\footnote{Recently collective-flow-like behaviors are reported in bulk hadron production at low momenta in p$A$ collisions~\cite{Abelev:2012ola,Abelev:2014mda,Aad:2012gla,Aad:2013fja,Aad:2014lta,CMS:2012qk,Chatrchyan:2013nka,Adare:2013piz,Adare:2014keg}.}. At forward rapidities at the LHC, the relevant $x$ values become very small ($\sim 10^{-5}$) even for the heavy quark production, and the corresponding saturation scale will be $Q_s^2\gtrsim 15~(\text{GeV}/c)^2$.
Therefore, besides light hadron productions, heavy quark production can be used as a quantitative probe for the gluon saturation in the target nucleus at the LHC.
At the same time, we note that p$A$ collisions provide an important baseline for quantifying the nuclear modification of hadron productions by a quark-gluon medium created in $AA$ collisions.

Open heavy flavor meson productions ($D$ and $B$) in pp and p$A$ collisions were previously evaluated in the CGC framework at mid rapidity at the LHC~\cite{Fujii:2013yja}. In the RHIC and LHC experiments, the decay leptons -- the electrons $e$ at mid and the muons $\mu$ at forward rapidity -- are also detected for the study of  heavy quark production. Importantly, the muon detection is the only observable relevant to the heavy flavor production at forward rapidity in the RHIC and LHC experiment setup. Then we need to evaluate the leptonic decays of open heavy flavor mesons ($Q\rightarrow Xl\nu$ channel) in order to study the saturation effects in the observed forward muon spectra. As was done in Refs.~\cite{Cacciari:2005rk,Luszczak:2008je}, we can compute the lepton production from the semileptonic decays of the heavy flavor mesons by convoluting the meson production cross-sections with the decay functions.
The main purpose of this study is to quantify in the CGC framework the sensitivity of the single lepton spectra to the small-$x$ gluon distribution in the target nucleus.

In the CGC framework, the heavy quark pair production cross-section is available analytically at the leading order in the strong coupling constant $\alpha_s$ and at the order ${\cal O}(\rho_{\rm p}^1\rho_A^\infty)$ for a dilute-dense system, where $\rho_{\rm p}$ ($\rho_A$) is the color charge density in the proton (nucleus). To this order, the physical picture of the pair production is simple: one is the process in which an incoming gluon from the proton splits into a quark pair and the pair is multiply scattered in the target nucleus, and the other is the process in which an incoming gluon is multiply scattered in the target nucleus and then splits into a quark pair.

The multiple scattering of the quark pair in the target in the eikonal approximation is represented by multipoint functions, which obey the so-called JIMWLK evolution equation in the rapidity $Y=\ln(1/x)$~\cite{Gelis:2010nm,Kovchegov:2012mbw}. In the large-$N_c$ approximation with $N_c$ being the number of colors, these general multipoint functions are written in terms of the two-point function (dipole amplitude) and the four-point function. Especially, the quark-pair production cross-section inclusive in the color degrees is written with the dipole amplitude only. The rapidity evolution of the dipole amplitude is controlled by the Balitsky-Kovchegov (BK) equation~\cite{Balitsky:1995ub,Kovchegov:1996ty} which sums up the small-$x$ quantum corrections of orders $(\alpha_sY)^n$ in the dilute regime and non-linear effects in the dense regime . At the present day, the BK equation with running coupling kernel (rcBK)~\cite{Balitsky:2006wa,Albacete:2007yr} is regarded as a useful tool for phenomenologies. The rcBK equation gives the appropriate evolution speed of the saturation scale $Q_s^2(x)$. The initial condition for the dipole amplitude in the rcBK equation is constrained by a global analysis of HERA DIS data.

This paper is organized as follows: In Sec.~2, we first give a brief review of heavy quark production in the CGC framework with the rcBK equation which is utilized in this paper. Then, we introduce a decay distribution function for describing the semileptonic decay of heavy mesons. After that, in Sec.~3, we show the results of single lepton production spectra and its nuclear modification factor.

\section{Framework}

\subsection{From heavy quark to heavy flavor meson}

\begin{figure}
\centering
 \includegraphics[height=3.5cm]{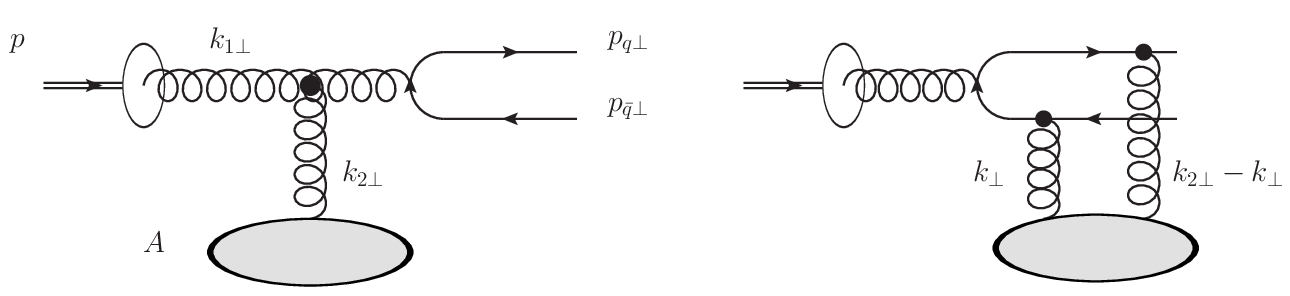} 
\caption{Leading order diagrams of heavy quark pair splitting before/after multiple scatterings with the nucleus.
Each blob represents multiple scatterings with the gluon fields in the target nucleus.}
\label{lo-diagrams}
\end{figure}

In the CGC framework, the heavy quark pair production cross-section 
at the leading order in $\alpha_s$ with the transverse momentum $p_{q\perp}~(p_{\bar q\perp})$ and rapidity $y_q~(y_{\bar q})$
of the  (anti-) quark in minimum bias events of a dilute-dense system (e.g., p$A$) is given in the large-$N_c$ limit as~\cite{Blaizot:2004wv}
\begin{align}
\frac{d \sigma_{q \bar{q}}}{d^2p_{q\perp} d^2p_{\bar{q}\perp} dy_q dy_{\bar{q}}}
=
\frac{\alpha_s^2}{64\pi^6 C_F}
\int\frac{d^2k_{2\perp}d^2k_\perp}{(2\pi)^4}
\frac{\Xi({\bs k}_{1\perp}, {\bs k}_{2\perp},{\bs k}_{\perp})}
{k_{1\perp}^2 k_{2\perp}^2}
\; 
\varphi_{{\rm p},x_1}(k_{1\perp})
\; 
\phi_{_A,x_2}^{q\bar{q},g}({\bs k}_{2\perp},{\bs k}_\perp)
\; ,
\label{eq:cross-section-LN}
\end{align}
where ${\bs p}_{q\perp} + {\bs p}_{\bar q \perp} = {\bs k}_{1\perp}+{\bs k}_{2\perp}$, 
$C_F = (N_c^2-1)/(2N_c)$, and $\varphi_{\rm p}$ and $\phi_A^{q\bar q,g}$ are the unintegrated gluon distribution function of the proton and the three-point function of the nucleus, respectively (see FIG.~\ref{lo-diagrams}). 
The values of Bjorken's $x$ on the proton and nucleus sides, $x_1$ and $x_2$,
respectively, are determined  in the $2 \rightarrow 2$ kinematics as 
$x_{1,2} = (m_{q\perp}/\sqrt{s_{NN}})e^{\pm y_{q}}+(m_{\bar q\perp}/\sqrt{s_{NN}})e^{\pm y_{\bar{q}}}$
with the transverse mass $m_{q\perp}=\sqrt{m^2+p_{q\perp}^2}$.\footnote{If we assume that the quark and anti-quark are produced with the same rapidity and transverse momentum, then $x_{1,2} = 2 (m_{q\perp}/\sqrt{s_{NN}}) e^{\pm y_{q}}$.}
The relevant hard matrix element $\Xi({\bs k}_{1\perp},{\bs k}_{2\perp}, {\bs k}_{\perp})$ can be found in Ref.~\cite{Fujii:2013gxa}.
The integrand depends on two transverse momenta, ${\bs k}_{2\perp}$ and ${\bs k}_{\perp}$, on the nucleus side, 
by which the $k_\perp$-factorization is broken in a strict sense~\cite{Fujii:2005vj}.
With this caution in mind, we call this formula $k_\perp$-factorized formula in a generalized version.

In the forward rapidity region, where $x_1$ becomes large while $x_2$ is much smaller,
we can use the hybrid formula~\cite{Fujii:2007ty,Fujii:2013gxa}:
\begin{align}
\frac{d\sigma_{q\bar{q}}}{d^2p_{q\perp} d^2p_{\bar{q}\perp} dy_{q} dy_{\bar q}}
=
\frac{\alpha_s^2}{16\pi^2 C_F}
\int \frac{d^2k_\perp}{(2\pi)^2}
\frac{\Xi_{\rm coll}({\bs k}_{2\perp},{\bs k}_{\perp})}{k_{2\perp}^2}
\;
 x_1 G(x_1,\mu)
\; 
\phi_{{A},x_2}^{q\bar{q},g}({\bs k}_{2\perp},{\bs k}_\perp)
\; ,
\label{eq:cross-section-LN-coll}
\end{align}
where $x_1 G$ is the collinear gluon distribution function with $\mu$ being factorization scale.
Since we work at the leading order in $\alpha_s$, $\mu$ here should be regarded as a model parameter.
The explicit expression for the hard matrix element $\Xi_{\rm coll}({\bs k}_{2\perp},{\bs k}_{\perp})$ 
after the collinear approximation done on the proton side can be found in Ref.~\cite{Fujii:2013gxa}.

The single heavy quark production cross-section is obtained by integrating the pair production cross-section over the phase space of the anti-quark:
\begin{align}
\frac{d \sigma_{q}}{d^2p_{q\perp}dy_q}=\int d^2p_{\bar{q}}dy_{\bar{q}}
\frac{d \sigma_{q \bar{q}}}{d^2p_{q\perp} d^2p_{\bar{q}\perp} dy_q dy_{\bar{q}}}.
\end{align}

The unintegrated gluon distribution function $\varphi_{{\rm p},x}({\bs k})$ of the proton is expressed
in terms of the Fourier transform
of the dipole amplitude $S_{Y}^{\rm adj}(k_{\perp})$ in the adjoint representation,
\begin{align}
\varphi_{{\rm p},x}(k_{\perp}) = \pi R_{\rm p}^2\, \frac{N_ck_{\perp}^2}{4\alpha_s} S_{Y}^{\rm adj}(k_{\perp})
\; ,
\end{align}
where $Y=\ln(x_0/x)$ is the evolution rapidity measured from $x_0$, and
the transverse geometry of the proton is replaced with the effective transverse area $\pi R_{\rm p}^2$.
In the large-$N_c$ limit, the adjoint dipole amplitude reduces to the square of the fundamental dipole
amplitude in the coordinate space, and therefore
\begin{align}
S_{Y}^{\rm adj}(k_{\perp})
= \int d^2x_\perp e^{ik_{\perp}\cdot x_\perp} 
{S}^2_{Y}(x_\perp)
=\int\frac{d^2l_\perp}{(2\pi)^2}S_{Y}({\bs k}_{\perp}-{\bs l}_\perp)S_{Y}(l_{\perp})
\end{align}
with 
\begin{align}
{S}_{Y}(k_{\perp})
\equiv\int d^2x_\perp e^{ik_{\perp}\cdot x_\perp} {S}_{Y}(x_\perp)
\; .
\end{align}
The collinear gluon distribution function of the proton $xG(x,\mu^2)$ is obtained by integrating $\varphi_{{\rm p},x}(k_{\perp})$ over the transverse momentum $k_\perp$ up to the factorization scale $\mu$:
\begin{align}
\frac{1}{4\pi^3}
\int^{\mu^2} d (k_{\perp}^2) \varphi_{{\rm p},x}(k_{\perp})
\equiv
xG(x,\mu^2).
\label{eq:xG-cgc}
\end{align}
We note here that $\varphi_{{\rm p},x}(k_{\perp})$ is different from the unintegrated gluon distribution function in a strict sense because the former includes higher twist effects. When we compare the gluon distribution obtained by Eq.~(\ref{eq:xG-cgc}) with the collinear gluon distribution, e.g., CTEQ6L PDF~\cite{Pumplin:2002vw}, we easily find a difference by a factor of two or so.

Similarly the three-point function of the nucleus can be expressed in the large-$N_c$ limit with the fundamental dipole amplitude  as
\begin{align}
\phi_{{A},x}^{q\bar{q},g}({\bs k}_{\perp},{\bs l}_\perp)
=\pi R_{A}^2 \,\frac{N_c k_{\perp}^2}{4\alpha_s}S_{Y}({\bs k}_{\perp}-{\bs l}_\perp)S_{Y}(l_\perp)
\; ,
\end{align}
where $\pi R_{A}^2$ is the effective transverse area of the target nucleus.
We ignore the transverse profile of the target nucleus here 
for simplicity. 
A recent attempt to include the centrality dependence in p$A$ collisions is found in Ref.~\cite{Lappi:2013zma}.
By construction the three-point function is related to the two-point function via
$\phi^{g,g}_{A,x}(k_{\perp})=\int \frac{d^2l_\perp}{(2\pi)^2}\phi_{{A},x}^{q\bar{q},g}({\bs k}_{\perp}, {\bs l}_\perp)$,
and this relation is respected in the large-$N_c$ approximation.

Rapidity dependence of the dipole amplitude $S_Y(r_\perp)$ in the fundamental representation 
is described in the large-$N_c$ limit by the rcBK equation~\cite{Balitsky:2006wa,Albacete:2007yr}. 
Using this solution, we express the three-point gluon function $\phi_{{A},x}^{q\bar{q},g}(k_{\perp},k_\perp)$ 
in the nucleus.
This equation resums quantum corrections of orders $(\alpha_s Y)^n$ in the linear regime, while by unitarity it tames the growth of the transition amplitude in the nonlinear regime.
Phenomenological analyses with the rcBK equation have been successful in describing HERA-DIS data as well as LHC hadron production data~\cite{Albacete:2012xq,Albacete:2014fwa}.

The rcBK equation is written in the coordinate space as
\begin{align}
-\frac{dS_{Y}({r_\perp})}{dY}
 = \int d^2 r_{1\perp} \mathcal{K}_{\rm Bal}(r_\perp, r_{1\perp}) 
\Big [  S_{Y}({r_\perp}) - S_{Y}({r_{1\perp}})S_{Y}({r_{2\perp}})
\Big ]
\; ,
\end{align}
where the evolution kernel in Balitsky's prescription~\cite{Balitsky:2006wa} is
\begin{align}
\mathcal{K}_{\rm Bal}(r_\perp,r_{1\perp})=&
\frac{\alpha_s (r^2) N_c} {2\pi^2}\,
\left [
\frac{1}{r_1^2} \left ( \frac{\alpha_s(r_1^2)}{\alpha_s(r_2^2)}-1  \right )
+
\frac{r^2}{r_1^2 r_2^2}
+
\frac{1}{r_2^2} \left ( \frac{\alpha_s(r_2^2)}{\alpha_s(r_1^2)}-1  \right )
\right ]
\label{eq:rcBK-kernel}
\end{align}
with ${\bs r}_\perp= {\bs r}_{1\perp}+ {\bs r}_{2\perp}$ being the transverse position of the dipole.
Here the running coupling constant in the coordinate space appears, for which
we assume the following form motivated by the one-loop expression in the momentum space:
\begin{align}
\alpha_s(r^2)= \left [\frac{9}{4\pi} \ln 
\left (\frac{4 C^2}{r^2 \Lambda^2}+a \right ) \right ]^{-1}
\, .
\end{align}
Note that we have introduced a smooth infrared cutoff $a$ so as to satisfy
$\alpha_s(r \to \infty)=\alpha_{\rm fr}$, as was used in Ref.~\cite{Fujii:2011fh}.

\begin{table}[t]
\renewcommand\arraystretch{1.2}
\begin{center}
\begin{tabular}{c||ccccc}
\hline
set & $Q_{s0,\rm p}^2/{\rm GeV}^2$ & $\gamma$ & $\alpha_{\rm fr}$ & $C$ & $e_c$\\
\hline 
MV$^\gamma$  & 0.1597 & 1.118 & 1.0 & 2.47 & 1\\
MV$^e$    & 0.06    & 1     & 0.7 & 2.68 & 18.9\\
\hline
\end{tabular}
\caption{Parameter values of the initial dipole amplitude and the running coupling constant in the coordinate space.
\label{tab:par}}
\end{center}
\end{table}

\begin{figure}
\centering
 \includegraphics[height=7.9cm,angle=270]{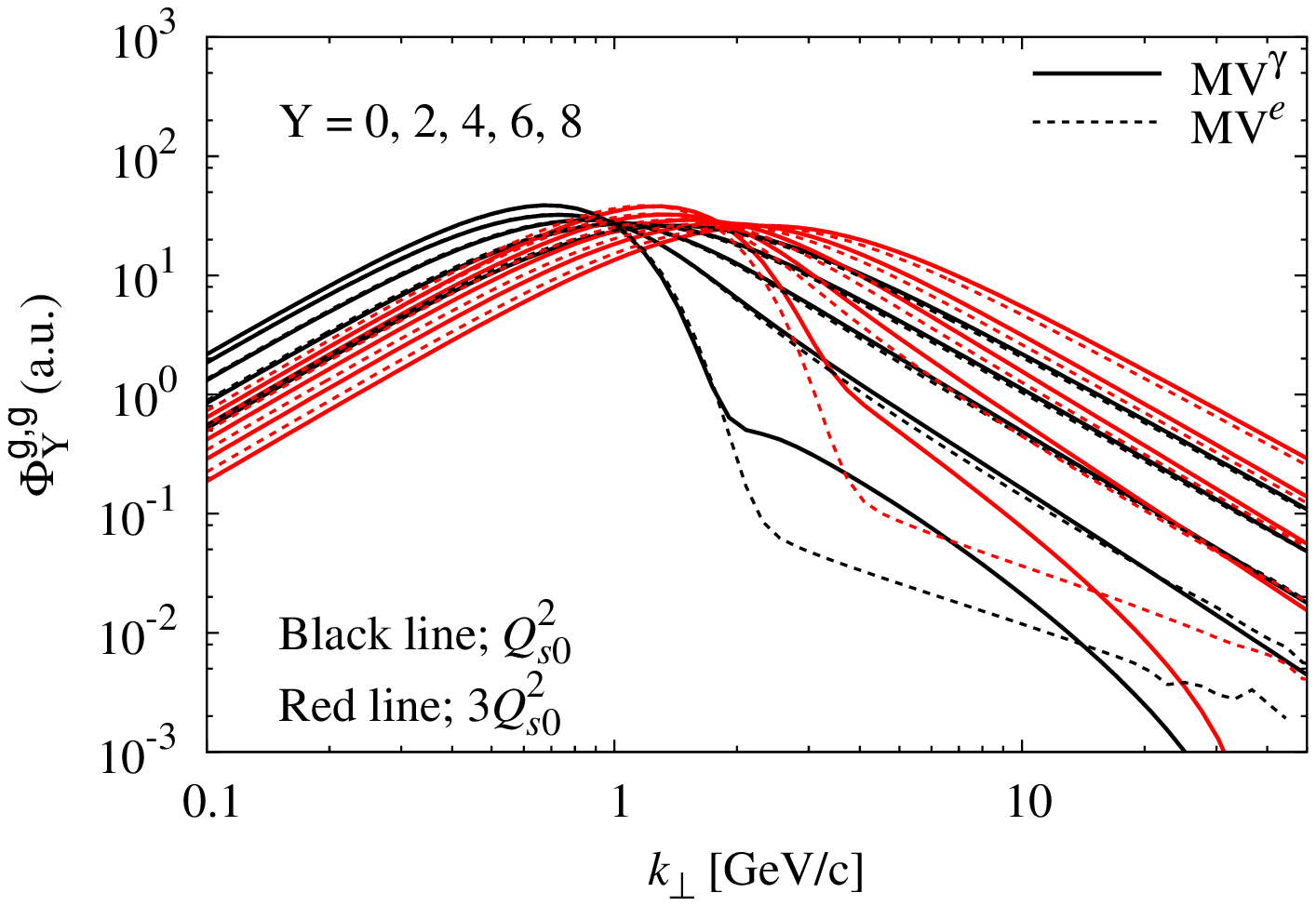} 
\caption{Gluon distribution functions $\Phi^{g,g}_{x}$ of MV$^\gamma$ (solid line) and MV$^e$ (dotted line).}
\label{phi_gg}
\end{figure}

By setting the initial condition for the rcBK equation at $x_0=0.01$
in a modified form from the McLerran-Venugopalan model~\cite{McLerran:1993ni}:  
\begin{align}
S_{{Y=0}}(r_\perp)=
\exp\left[-\frac{\left(r_\perp^2Q_{s0,p}^2\right)^\gamma}{4}\ln\left(\frac{1}{r_\perp\Lambda}+e_c\cdot e\right)\right]
\, ,
\end{align}
the dipole amplitude is constrained by global fit of HERA DIS data. 
As we have shown in Ref.~\cite{Fujii:2013yja}, the original MV model ($\gamma=1$) gives rise to a too-hard $p_\perp$ spectrum of heavy mesons at high $p_\perp$.
In this study we apply two parameter sets, MV$^{\gamma}$ and MV$^e$, as listed in Table~\ref{tab:par}~\cite{Fujii:2011fh,Lappi:2013zma} with $\Lambda=0.241$ GeV.
For a heavy nucleus we replace the initial saturation scale by $Q_{s0,A}^2=cA^{1/3}Q_{s0,p}^2$ with $c=0.5$ in this study, as suggested for minimum bias events
from nuclear DIS analysis in Ref.~\cite{Dusling:2009ni}.
In FIG.~\ref{phi_gg} we show the unintegrated gluon distributions of the proton and the nucleus for the sets, MV$^\gamma$ and MV$^e$.
We find that the difference between MV$^\gamma$ and MV$^e$ is very small in low $k_\perp\lesssim1~\text{GeV}$ region,
and that the difference seen in higher $k_\perp$ region diminishes with the rapidity evolution.
At larger $x \geq x_0$, 
we apply the following extrapolation ansatz~\cite{Fujii:2006ab,Fujii:2013gxa}:
$\phi_{A,x}^{q\bar{q},g}(l_\perp,k_\perp)=\phi_{A,x_0}^{q\bar{q},g}(l_\perp,k_\perp)
\left(\frac{1-x}{1-x_0}\right)^4 \left(\frac{x_0}{x}\right)^{0.15}$.

Production of heavy flavor meson $h$ ($D^0$, $D^+$, etc.) from the quark $q$ is described with the heavy-quark fragmentation function $D_q^h(z)$ as 
\begin{align}
\frac{d \sigma_{h}}{d^2p_{h\perp} dy}
=
Br(q\rightarrow h)
\int dz\frac{D_q^h(z)}{z^2}
\frac{d \sigma_{q}}{d^2p_{q\perp} dy}
\, ,
\label{eq:open-cross-section-LN}
\end{align}
where the rapidity is set to $y_q=y_{h}=y$ and the momentum fraction $z$ is defined by $p_{h\perp}=zp_{q\perp}$.
In this paper we use the fragmentation function of Kartvelishvili's form~\cite{Kartvelishvili:1977pi}
\begin{align}
D_q^h(z)=(\alpha+1)(\alpha+2) z^{\alpha}(1-z)
\end{align}
with the parameter $\alpha=3.5$ $(13.5)$ for $D$ $(B)$ meson.
We expect that the discussion below will be almost unchanged when we use another form for $D(z)$.
The branching ratio $Br(q\rightarrow h)$ for the quark ($q$) to make a transition to the heavy flavor meson $h$
satisfies $\sum_h Br(q\rightarrow h)=1$.
For example, $Br(c\rightarrow D^0)=0.565$~\cite{PDG2008}.

\subsection{From heavy flavor meson to lepton}

Now we proceed to lepton production from semileptonic decays of heavy flavor mesons $h \to X l \bar \nu$,
which can be computed by convoluting the meson production cross-section with the lepton decay function $\cal F$. Then the cross-section is given with the use of Eq.~(\ref{eq:open-cross-section-LN}) as:
\begin{align}
\frac{d \sigma_{l}}{d^2p_{l\perp} dy_l}
=
\int
dp_{h\perp}p_{h\perp}dy_h~{\cal F}(p_{l},p_{h})~
\frac{d \sigma_{h}}{d^2p_{h\perp} dy_h}.
\;
\label{eq:lepton-xsection}
\end{align}
Here ${\cal F}(p_{h},p_{l})$ is the probability for the lepton with momentum $p_{l}$ to be
produced in the decay of the heavy meson with momentum $p_h$ in the laboratory frame,
and is expressed as an integral over the phase space~\cite{Gronau:1976ng,Ali:1977eu}:
\begin{align}
{\cal F}(p_{h},p_{l})=\int d\phi\frac{M_h}{4\pi (p_{h}\cdot p_{l})}
f\left(\frac{ p_{h}\cdot p_{l}}{M_h}\right)
\, ,
\label{eq:angular-integral}
\end{align}
where $\phi$ is the azimuthal angle between $p_{h\perp}$ and $p_{l\perp}$. 
$f(E_l)$ is the distribution of the lepton with energy $E_l$ in the heavy-meson rest frame 
and is parametrized\footnote{An update of the lepton decay function is available in Ref.~\cite{Luszczak:2008je} where they fit more recent experimental data.} as 
\begin{align}
f(E_l)&=\omega\frac{E_l^2(M_h^2-M_X^2-2M_hE_l)^2}{M_h-2E_l}.
\label{eq:decay-distribution}
\end{align}
The normalization factor $\omega = 96/ [(1-8t^2+8t^6-t^8-24t^4\ln t)M_h^6]$ with $t = M_X / M_h$. 
The mass $M_X$ of the produced particle $X$ is set to 
$M_X=M_K=0.497$ GeV in the $D$ decay ($M_h=M_D=1.86$ GeV)
and $M_X=M_D=1.86$ GeV in the $B$ decay ($M_h=M_B=5.28$ GeV). 
We neglect the lepton masses here because $m_e/m_h,  m_\mu/m_h\ll 1$.

\section{Heavy-quark decay leptons}

Here we show the numerical results of the particle spectra computed at RHIC and LHC energies. 
We adopt CTEQ6L PDF~\cite{Pumplin:2002vw} with $\mu=\sqrt{m^2+p_\perp^2}$
for the collinear gluon distribution function  in the hybrid model. 
Here we neglect the contribution of $b\rightarrow c\rightarrow l$ channel because we expect that this channel gives only a small contribution 
compared to $c\rightarrow l$ channel at low $p_\perp$ and $b\rightarrow l$ at high $p_\perp$.
Through this paper, we set $R_{\rm p}=0.8$ fm for the proton's transverse radius and fix $\alpha_s=0.2$ in front of the hard matrix elements
and use the following branching ratios for the decays of the quark to the lepton~\cite{Cacciari:2005rk}: $Br(b\rightarrow l)=0.1086$ for $B$ decay and $Br(c\rightarrow l)=0.103$ for $D$ decay.

\subsection{pp collisions}

The quark-pair production formula Eq.~(\ref{eq:cross-section-LN}) is derived for asymmetric collision systems, and therefore is not best suitable for pp collisions at mid rapidity. Nonetheless it appears as a reference in calculation of the nuclear modification factor at mid rapidity. Let us examine the lepton production in pp collisions here.

\begin{figure}[tbp]
 \centering
 \includegraphics[height=7.9cm,angle=270]{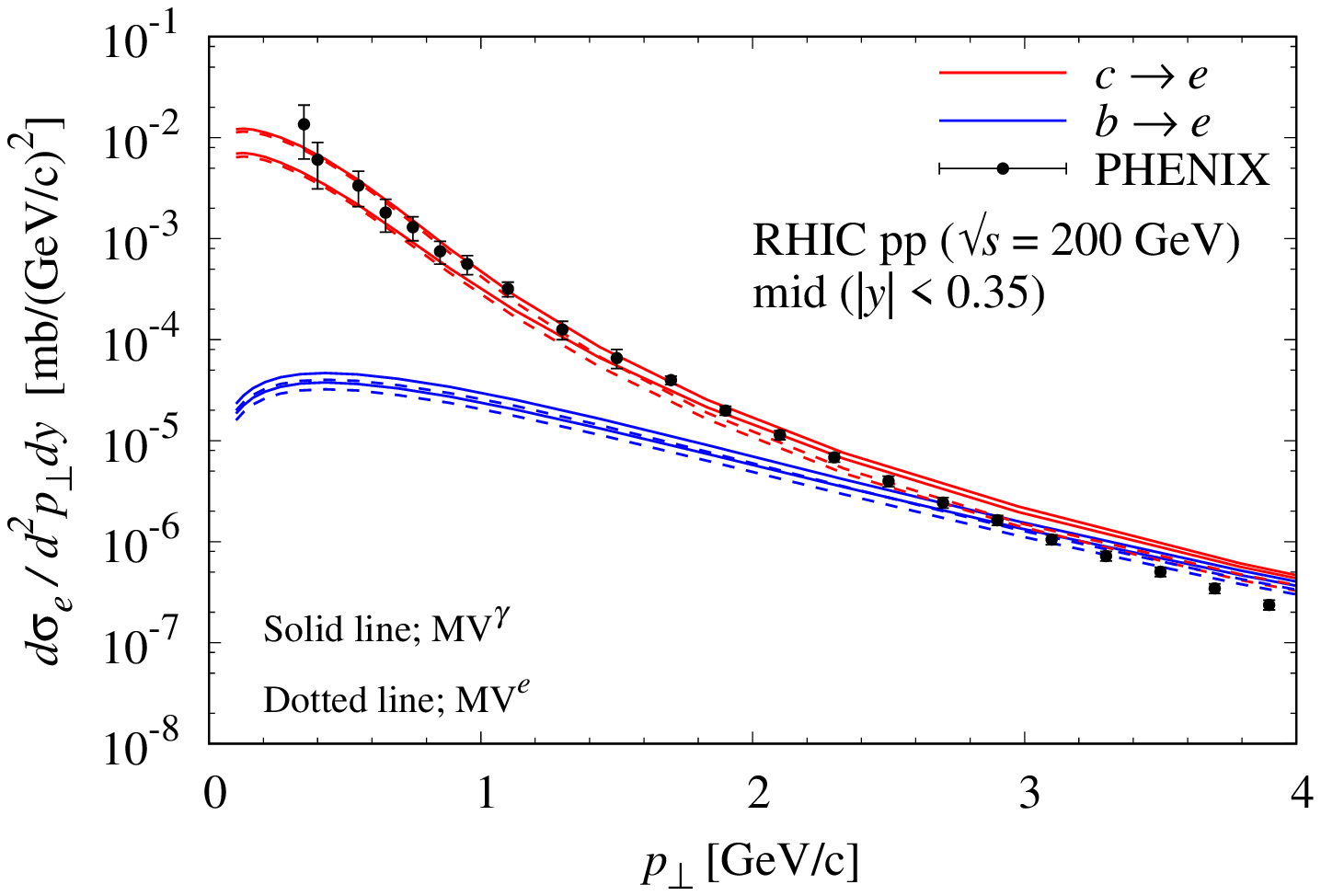} 
 \includegraphics[height=7.9cm,angle=270]{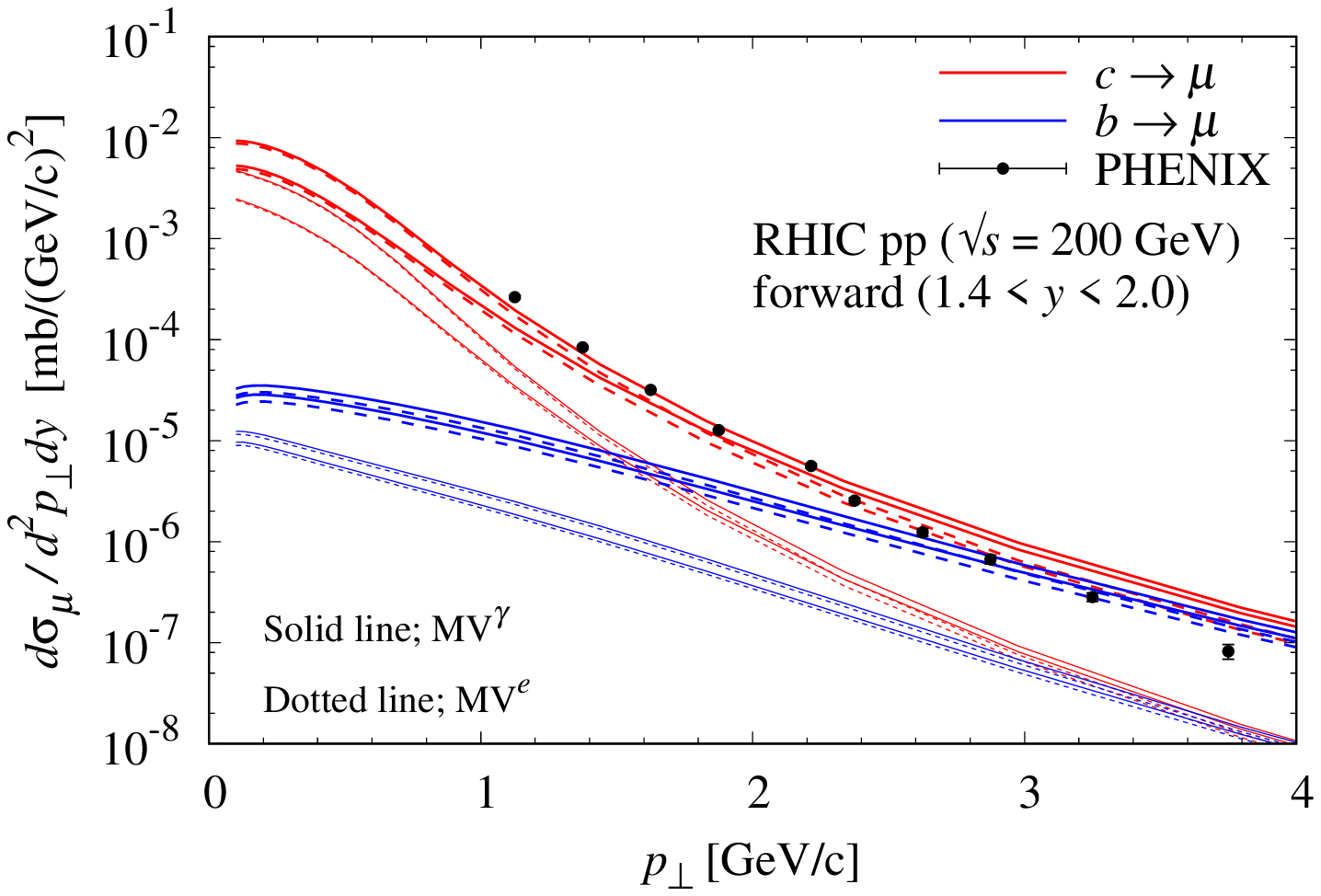}
 \caption[*]{Double differential cross-section of the heavy-quark decay electrons (left) and muons (right) as a function of transverse momentum $p_\perp$ at $|y|<0.35$ and $1.4<y<2.0$, respectively, in pp collisions at $\sqrt{s_{NN}}=200$ GeV. Red (blue) thick solid lines show the lepton yields from the charm (bottom) decays obtained by Eq.~(\ref{eq:cross-section-LN}) with MV$^\gamma$. The uncertainty band represents the change of quark mass scale: $m_c=1.2\sim1.5$ GeV for charm and $m_b=4.5\sim4.8$ GeV for bottom. Dotted lines denote the results with MV$^e$. 
The results obtained by Eq.~(\ref{eq:cross-section-LN-coll}) are depicted with thin lines.
RHIC data at mid rapidity is taken from Ref.~\cite{Adare:2010de} and forward data is taken from~\cite{Adare:2013lkk}.
 }
\label{results-rhic-pp}
\end{figure}

FIG.~\ref{results-rhic-pp} shows the $p_\perp$ distribution of leptons from charmed-hadron and bottomed-hadron semileptonic decays in pp collisions at RHIC.
In the left panel, we see that at mid rapidity the CGC result for the charm-decay electrons (red solid line) gives a slight harder $p_\perp$ slope compared with the data.
The cross-section of the electrons from bottomed-hadron decay is much smaller than that from charmed-hadron decay at lower $p_\perp$.
Thus we neglect the bottomed-hadron decay contribution as long as we focus on the lepton production less than $p_\perp\sim 2$ GeV.

We remind here that, at mid rapidity at RHIC energy $\sqrt{s_{NN}}=200$ GeV, both $x_1$ and $x_2$ are larger than $x_0=0.01$ in heavy quark pair production, and that the CGC framework may be only marginally applicable in this kinematical region. Especially, 
the lepton yields solely depend on the initial condition for the gluon distributions and their extrapolation to larger $x>x_0$.
In the left panel of FIG.~\ref{results-rhic-pp} we find little difference between the results with MV$^\gamma$ (solid) and MV$^e$ (dotted) at $p_\perp\gtrsim 1$GeV.

At forward rapidities, $x_2$ ($x_1$) becomes smaller (larger) than $x_0$ and there appears some room for the effects of the quantum evolution \'a la rcBK to show up in the $p_\perp$ spectra. 
In the right panel of FIG.~\ref{results-rhic-pp}, we find that the lepton spectrum is still harder than the data at $1.4<y<2.0$. The difference between the results with the parametrizations MV$^\gamma$ and MV$^e$ is very weak. In this asymmetric case, we apply the hybrid formula. The resultant $p_\perp$ spectrum of the muon from charm decay is smaller than the data.

\begin{figure}[tbp]
 \centering
 \includegraphics[height=7.9cm,angle=270]{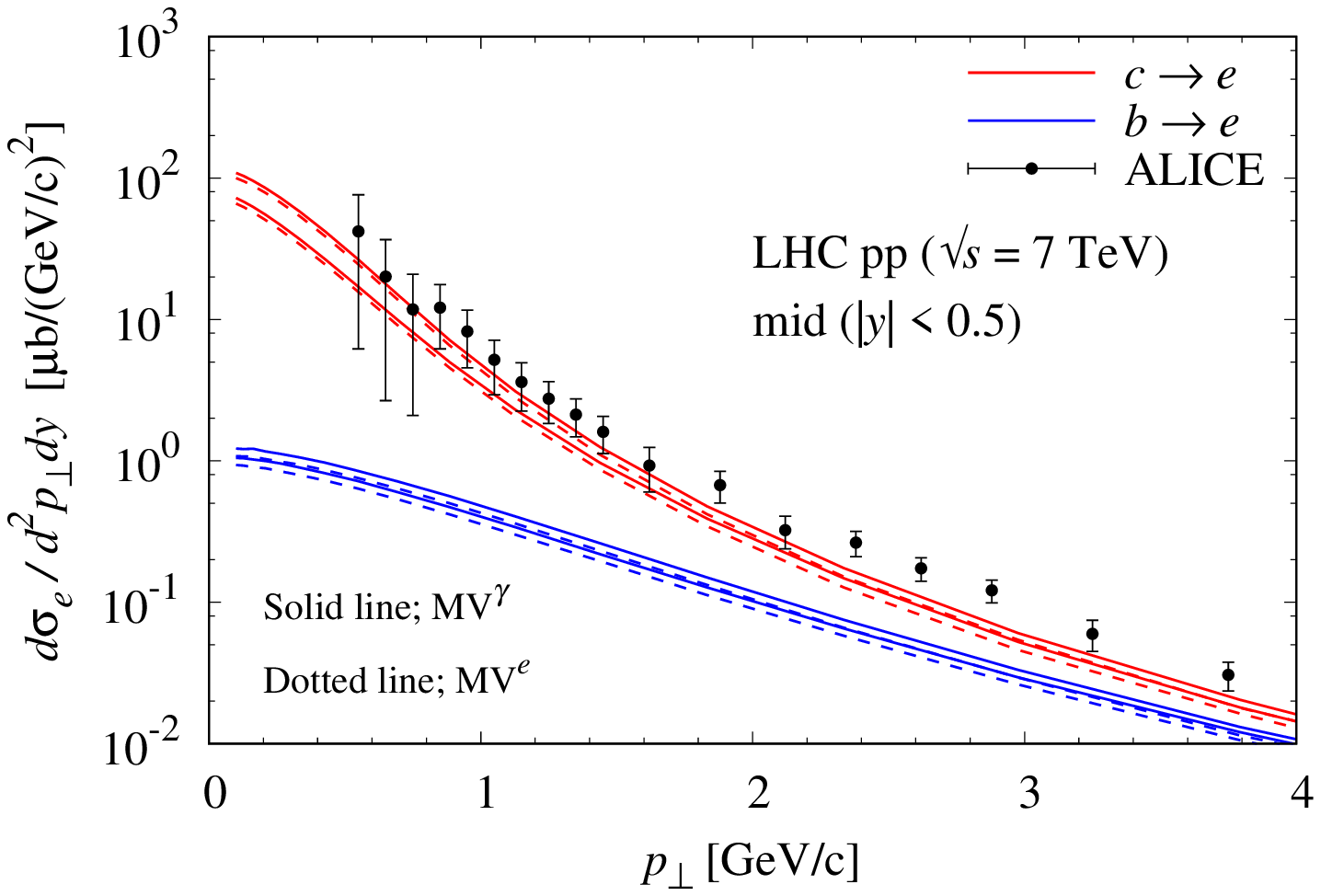} 
 \includegraphics[height=7.9cm,angle=270]{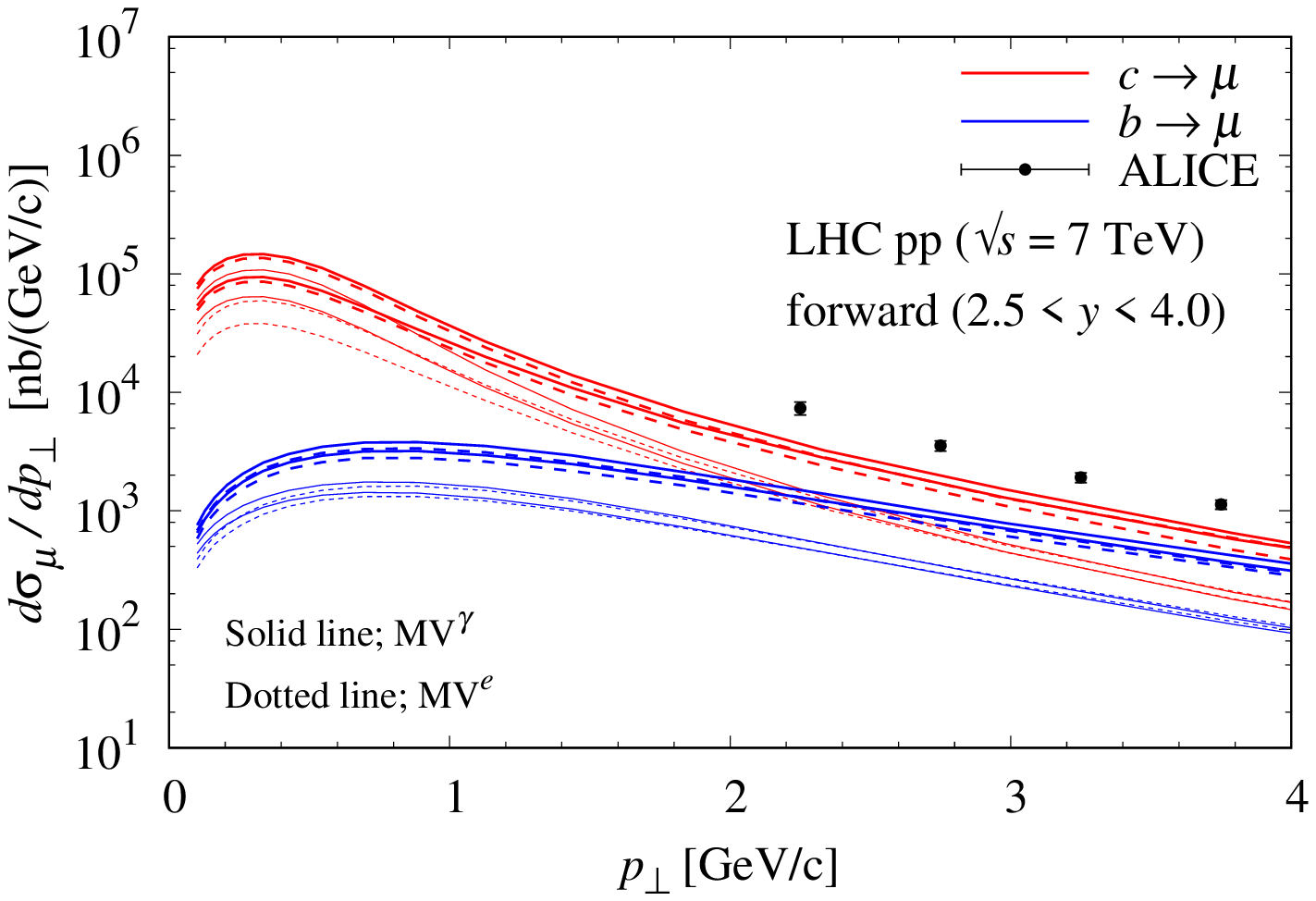}
 \caption[*]{Double differential cross-section of the heavy-quark decay electrons (left) and muons (right) from heavy flavor quark decays as a function of transverse momentum $p_\perp$ at $|y|<0.5$ and $2.5<y<4.0$, respectively, in pp collisions at $\sqrt{s_{NN}}=7$ TeV. 
 Notations are the same as in FIG.~\ref{results-rhic-pp}.
 LHC data at mid rapidity is taken from Ref.~\cite{Abelev:2012xe} and forward data is taken from~\cite{Abelev:2012pi}.
 }
 \label{results-lhc-pp}
\end{figure}

At the LHC energy $\sqrt{s_{NN}}=7$ TeV, the $x$ values become much smaller and the use of the CGC framework can be more justified.
The lepton spectra at mid (electron) and forward (muon) rapidities are 
shown in FIG.~\ref{results-lhc-pp}. The difference between the numerical results and the data becomes slightly smaller.
The hybrid formula gives rise to more or less a reasonable description for the $p_\perp$ slope of the muon at forward rapidity both with MV$^\gamma$ and MV$^e$, although the $p_\perp$ spectrum of the muon is smaller than the data.
The leptons from the b decays (shown in blue lines) are much suppressed as compared to those from the c decays in this momentum region $p_\perp \lesssim 4$ GeV.

\subsection{p$A$ collisions and nuclear modification factor}

\begin{figure}[tbp]
 \centering
 \includegraphics[height=7.9cm,angle=270]{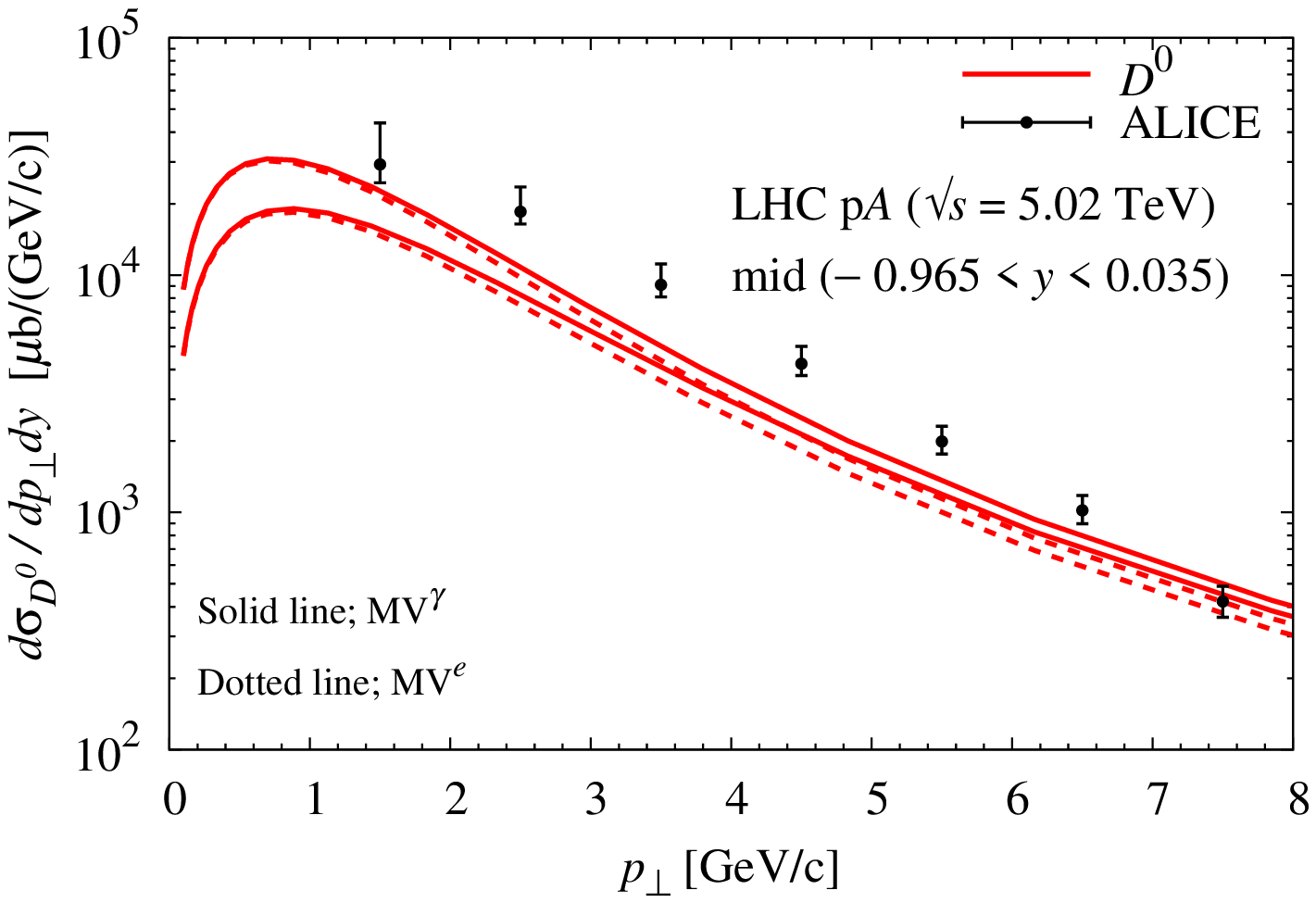} 
 \includegraphics[height=7.9cm,angle=270]{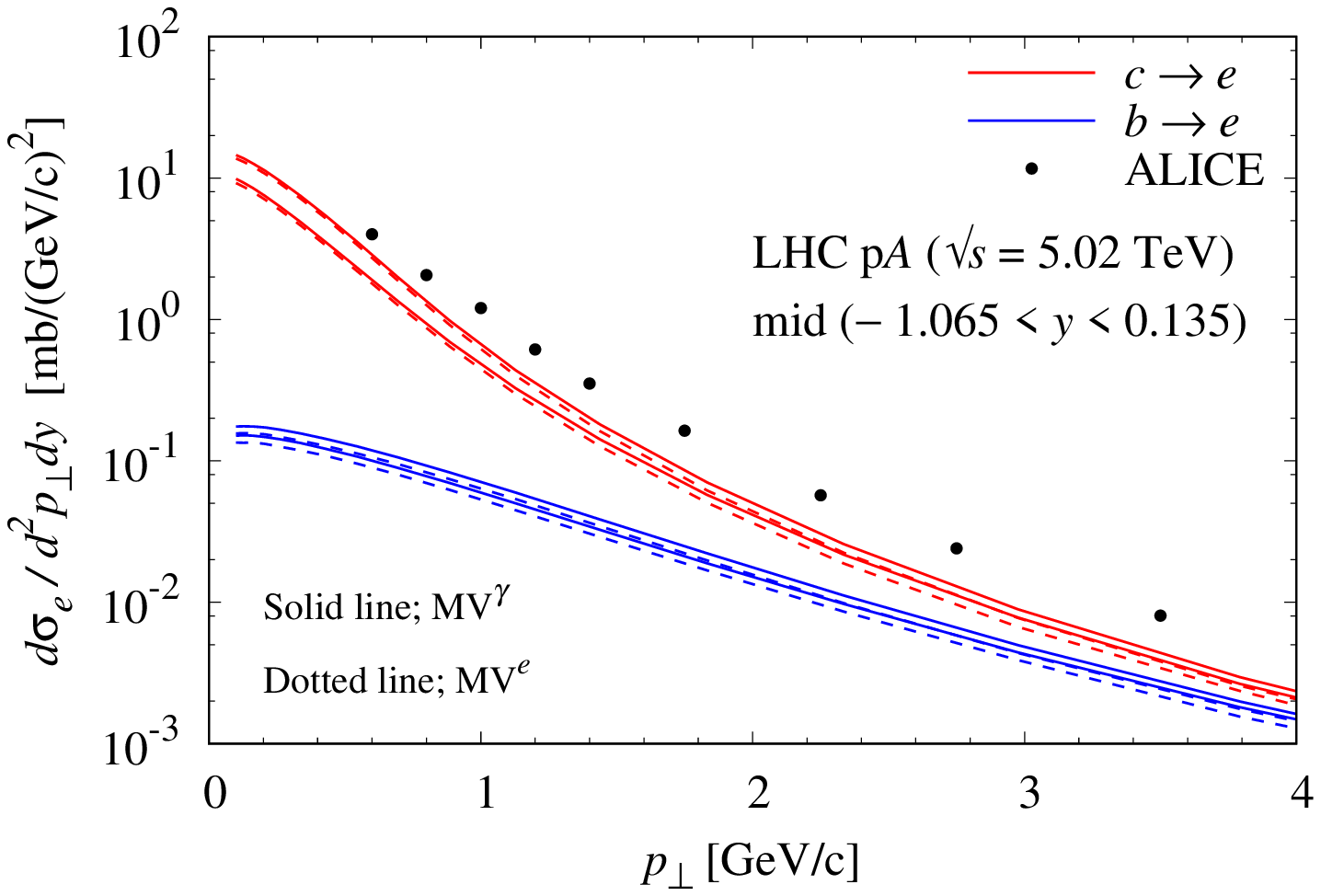} 
 \caption[*]{Left: Double differential cross-section of $D^0$ meson as a function of transverse momentum $p_\perp$ at $-0.965 < y < 0.035$ in pPb collisions at $\sqrt {s_{NN}}=5.02$ TeV. 
 Right: Double differential cross-sections of electron from charmed-hadron decay and bottomed-hadron decay as a function of $p_\perp$ at $-1.065 < y < 0.135$ in pPb collisions at $\sqrt{s_{NN}}=5.02$ TeV. 
Notations are the same as in FIG.~\ref{results-rhic-pp}.
LHC data for $D$ meson production is taken from Ref.~\cite{Abelev:2014hha} and for electron from Ref.~\cite{Adam:2015qda}.
 }
 \label{results-lhc-pA}
\end{figure}

Next, let us turn to heavy-quark decay leptons in p$A$ collisions. 
We have now two parameters to be fixed for the heavy nucleus,
the initial saturation scale for the rcBK evolution and the transverse size of the nucleus.
Regarding the initial saturation scale, we choose $Q_{s0,A}^2=3Q_{s0,{\rm p}}^2$ for the heavy nuclei, Au and Pb.~\footnote{This saturation scale is smaller than the naive expectation $Q_{s0,A}^2 = A^{1/3}Q_{s0,{\rm p}}^2$, which we employed in \cite{Fujii:2013gxa,Fujii:2013yja}. Ref.~\cite{Ma:2015sia} used $Q_{s0,A}^2\sim(2-3)Q_{s0,{\rm p}}^2$ for heavy quark pair production.}
As to the transverse radius of the nucleus, we determine it by 
imposing the nuclear modification factor $R_{{\rm p}A}=1$ at high $p_\perp$.

The nuclear modification factor is defined by
\begin{align}
R_{{\rm p}A}(p_\perp)=\frac{1}{A}\frac{d^3\sigma_{{\rm p}A}/d^2p_\perp dy}{d^3\sigma_{\rm pp}/d^2p_\perp dy}
\end{align}
and it is expected to scale as
$R_{{\rm p}A}\sim\frac{1}{A}\frac{\pi R_A^2Q_{sA}^{2\gamma}}{\pi R_{\rm p}^2Q_{sp}^{2\gamma}}$ 
for large $p_\perp$ in the CGC formula.
In order for $R_{{\rm p}A}=1$ at high $p_\perp$, we choose here the nuclear radius to be
$R_A=\sqrt{\frac{A}{(cA^{1/3})^\gamma}}R_{\rm p}\sim \sqrt{\frac{A}{3^\gamma}}R_{\rm p}$ as is done in Ref.~\cite{Ma:2015sia}.
This condition with parametrization MV$^\gamma$ (MV$^e$) gives $R_A=6.24$ (6.66) fm for Pb and 6.08 (6.48) fm for Au.
Our calculation can predict now only the $p_\perp$ and $y$ dependence of the nuclear modification factor $R_{{\rm p}A,y}(p_\perp)$, which is constrained to be 
$R_{{\rm p}A,y}(p_\perp)=1$ at high $p_\perp$.

\begin{figure}[tbp]
 \centering
 \includegraphics[height=7.9cm,angle=270]{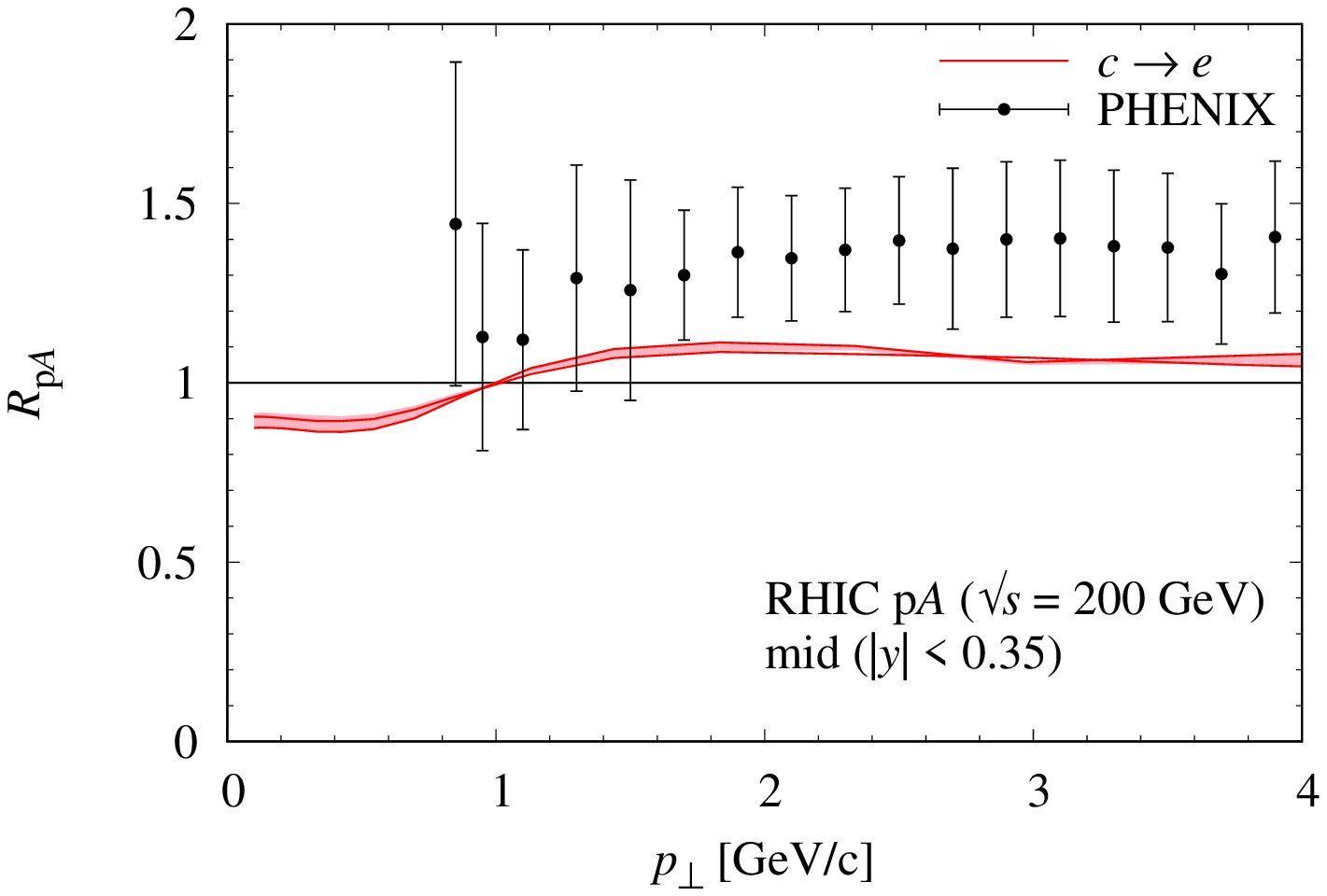} 
 \includegraphics[height=7.9cm,angle=270]{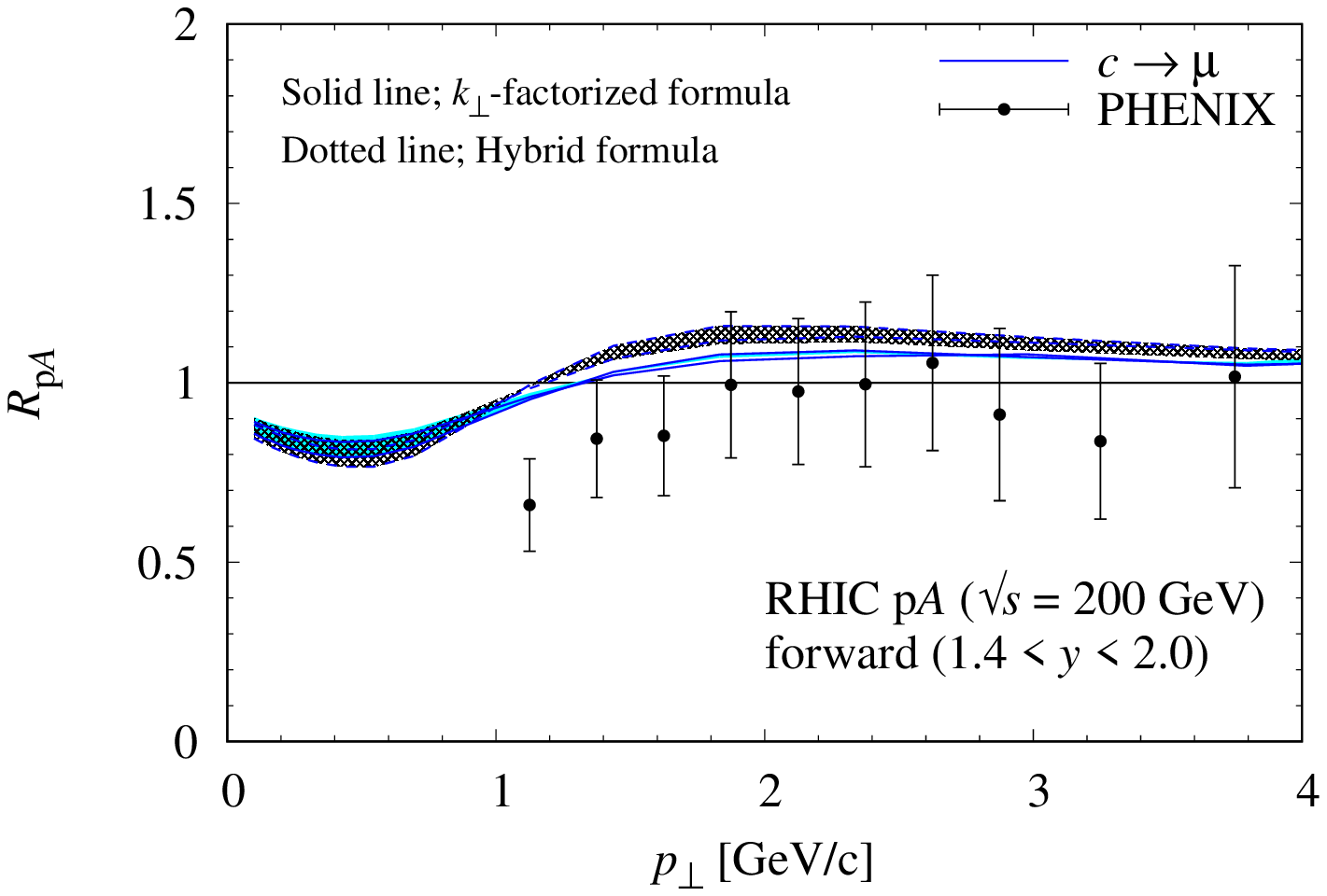} 
 \caption[*]{Nuclear modification factor at $\sqrt{s_{NN}}=200$ GeV at mid and forward rapidity.
 Solid (dotted) lines are obtained by using Eq.~(\ref{eq:cross-section-LN}) (Eq.~(\ref{eq:cross-section-LN-coll})). The quark mass scale dependence is shown by the different line. 
 Filled and Shaded bands are the uncertainty which is reflecting the difference of the initial condition between MV$^\gamma$ and MV$^e$ for Eq.~(\ref{eq:cross-section-LN}) and (\ref{eq:cross-section-LN-coll}), respectively.
RHIC data at mid rapidity is taken from Ref.~\cite{Adare:2012yxa} and forward data is taken from~\cite{Adare:2013lkk}.
 }
 \label{rpa-rhic}
\end{figure}

\begin{figure}[tbp]
 \centering
 \includegraphics[height=7.9cm,angle=270]{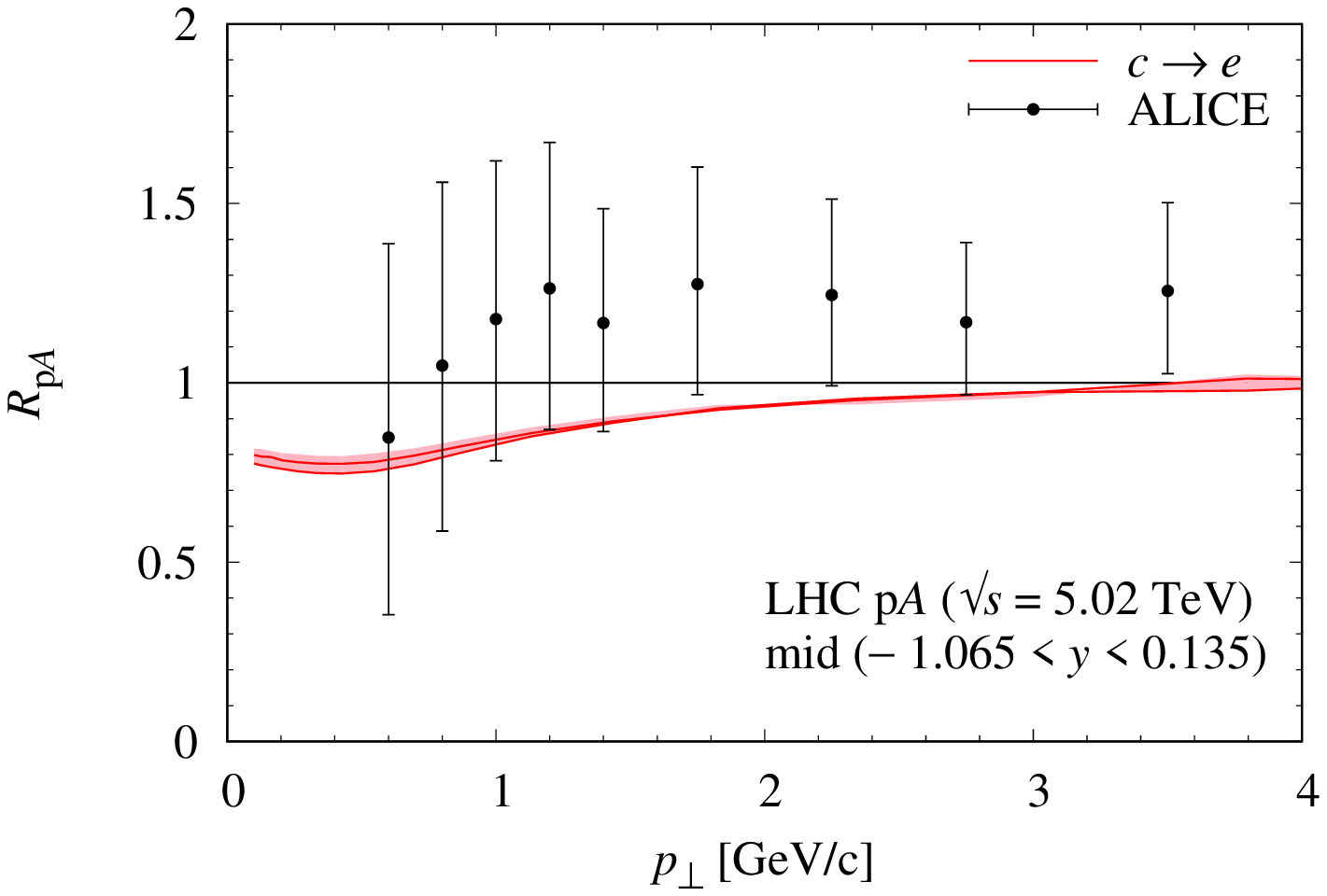} 
 \includegraphics[height=7.9cm,angle=270]{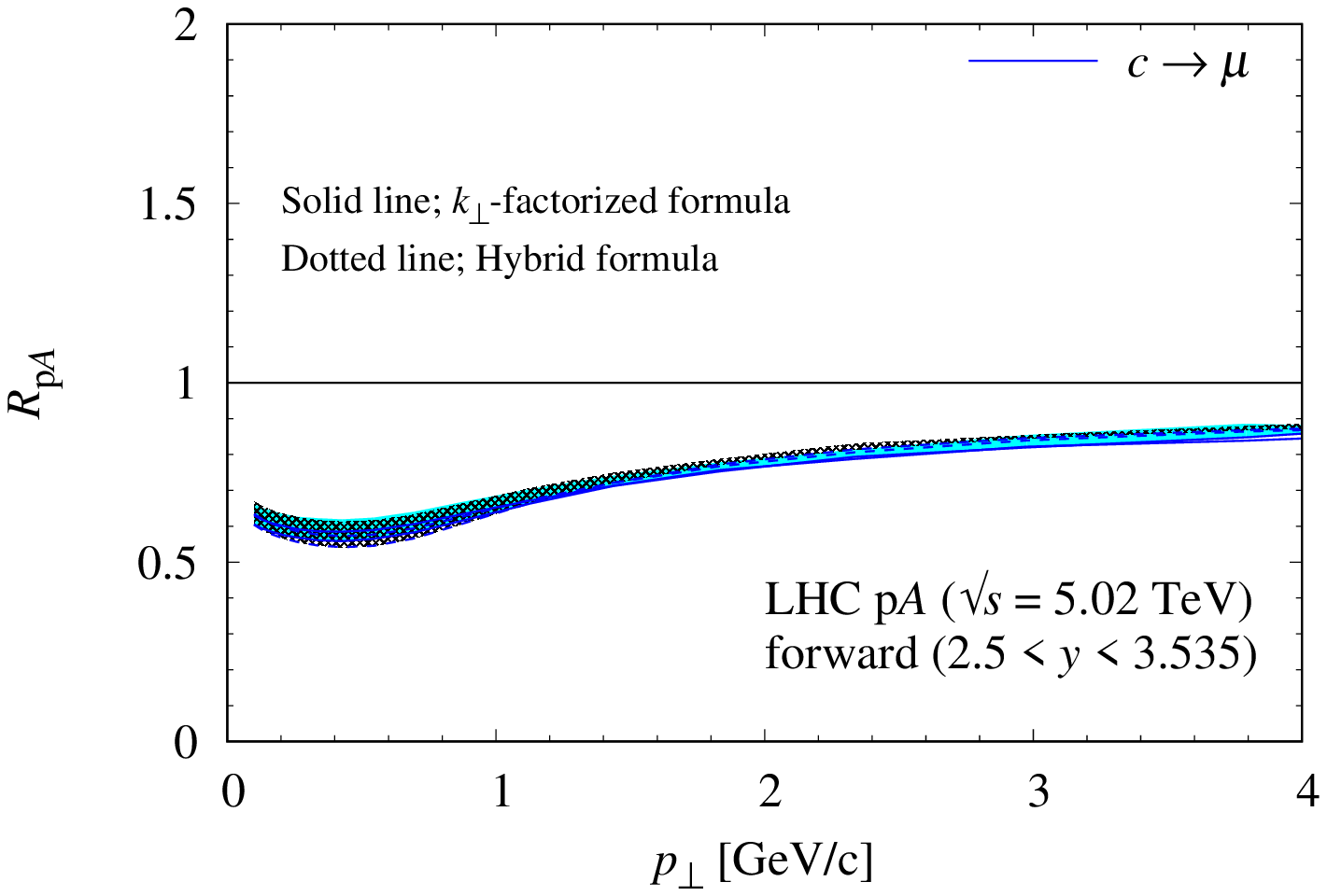} 
 \caption[*]{Nuclear modification factor at $\sqrt{s_{NN}}=5.02$ TeV.
 Notations are the same as in FIG.~\ref{rpa-rhic}.
LHC data at mid rapidity is taken from Ref.~\cite{Adam:2015qda}
 }
 \label{rpa-lhc}
\end{figure}

We use Eq.~(\ref{eq:cross-section-LN}) to evaluate  
the $p_\perp$ spectra of the $D^0$ and the electron at mid rapidities in p$A$ collisions at the LHC energy $\sqrt{s_{NN}}=5.02$ TeV, which is presented 
in FIG.~\ref{results-lhc-pA}. 
We find that the electron $p_\perp$ spectrum for charmed-hadron decay of our numerical result is slightly smaller than the data.
Note that the normalization is fixed by the condition for $R_{{\rm p}A}$ to become unity at large $p_\perp$.

Now we can compute $R_{{\rm p}A}$ of the leptons from charmed-hadron decays. 
We expect that the systematic uncertainties of the calculation partially cancel out 
in the ratio.
In FIG.~\ref{rpa-rhic} we show the results of $R_{{\rm p}A}$ of the electrons at mid rapidities $|y|<0.35$ and the muons at forward rapidities $1.4<y<2.0$ at $\sqrt{s_{NN}}=200$ GeV, 
together with the RHIC data. 
At mid rapidity, we find that $R_{{\rm p}A}$ is almost flat with weak suppression at $p_\perp\lesssim1$ GeV and subtle enhancement at $p_\perp\sim2$ GeV.
This structure stems from the multiple scattering effects encoded in the dipole amplitude, while
the evolution at $x_2<x_0$ is inoperative in the charm production here.
The experimental data of $R_{{\rm p}A}$ seems systematically larger than unity with some uncertainties.
At forward rapidity, where the charm production gets sensitivity to the gluon distribution at $x_2<x_0$, the $p_\perp$ dependence of $R_{{\rm p}A}$ for the muons becomes slightly more noticeable due to the quantum evolution. The data is now almost consistent with unity.
We also note that the $k_\perp$ factorized formula (\ref{eq:cross-section-LN}) and the hybrid formula (\ref{eq:cross-section-LN-coll}) give the same result.

In FIG.~\ref{rpa-lhc} we show the electron $R_{{\rm p}A}$ computed with Eq.~(\ref{eq:cross-section-LN}) at mid rapidity $-1.065<y<0.135$ at $\sqrt{s_{NN}}=5.02$ TeV.
It is suppressed at low $p_\perp$ and is recovering to unity monotonically with increasing $p_\perp$.This behavior reflects the $x$-evolution of the gluon distribution, but the smearing through the fragmentation and decay processes seems to make the effect relatively less prominent (see FIG.~\ref{D-rpa-lhc} below for $D$ meson case).
We also show in FIG.~\ref{rpa-lhc} a model prediction of $R_{{\rm p}A}$ at forward rapidity
at $\sqrt{s_{NN}}=5.02$ TeV.
As is found in $D$ meson and $J/\psi$ productions~\cite{Fujii:2013gxa,Fujii:2013yja}, 
we also expect a stronger suppression of the lepton $R_{{\rm p}A}$ at forward rapidity compared to that of at mid rapidity due to the stronger saturation effect.

\subsection{Contributions from saturation and extended scaling regions}

\begin{figure}
 \centering
 \includegraphics[height=7.9cm,angle=270]{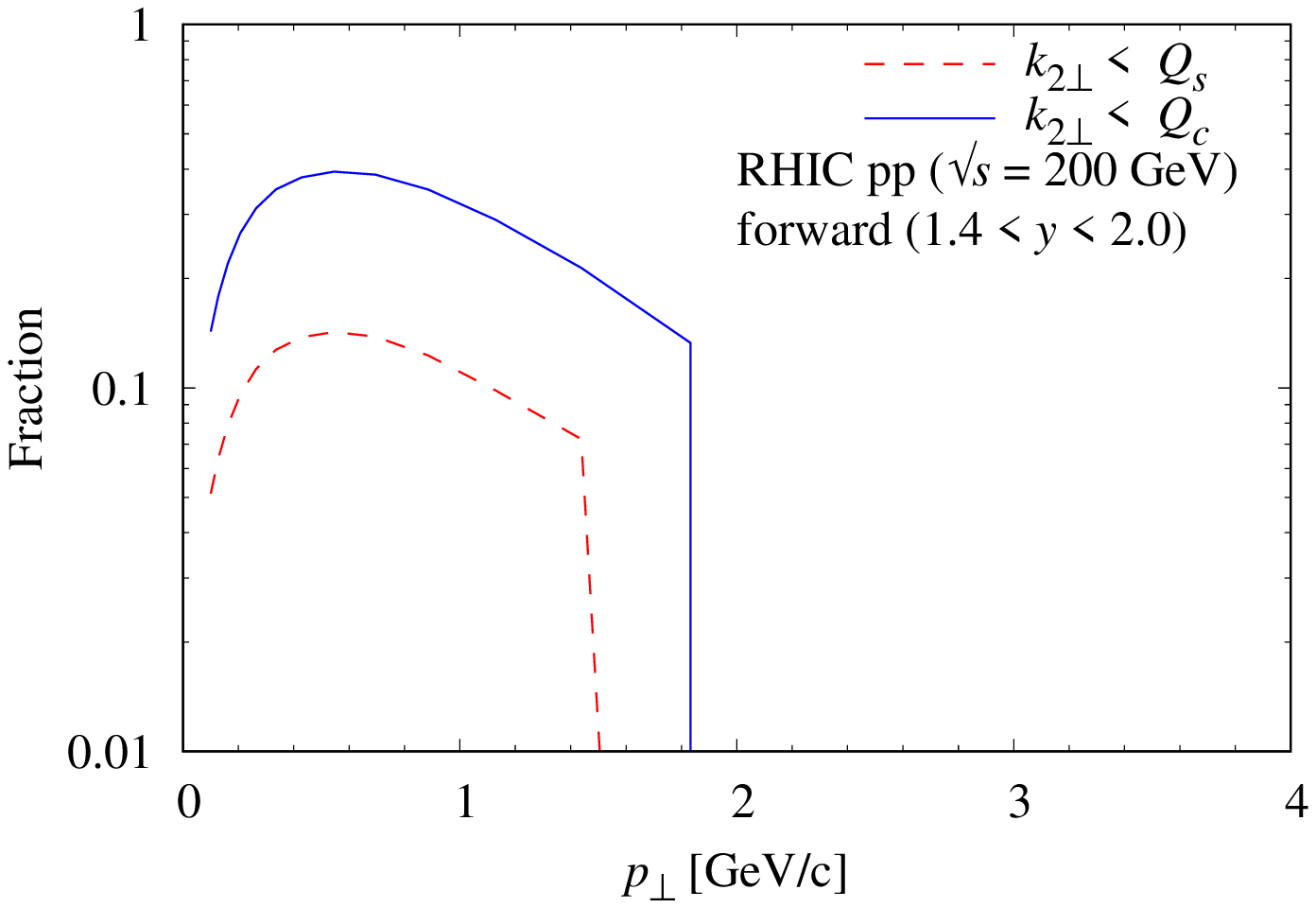} 
 \includegraphics[height=7.9cm,angle=270]{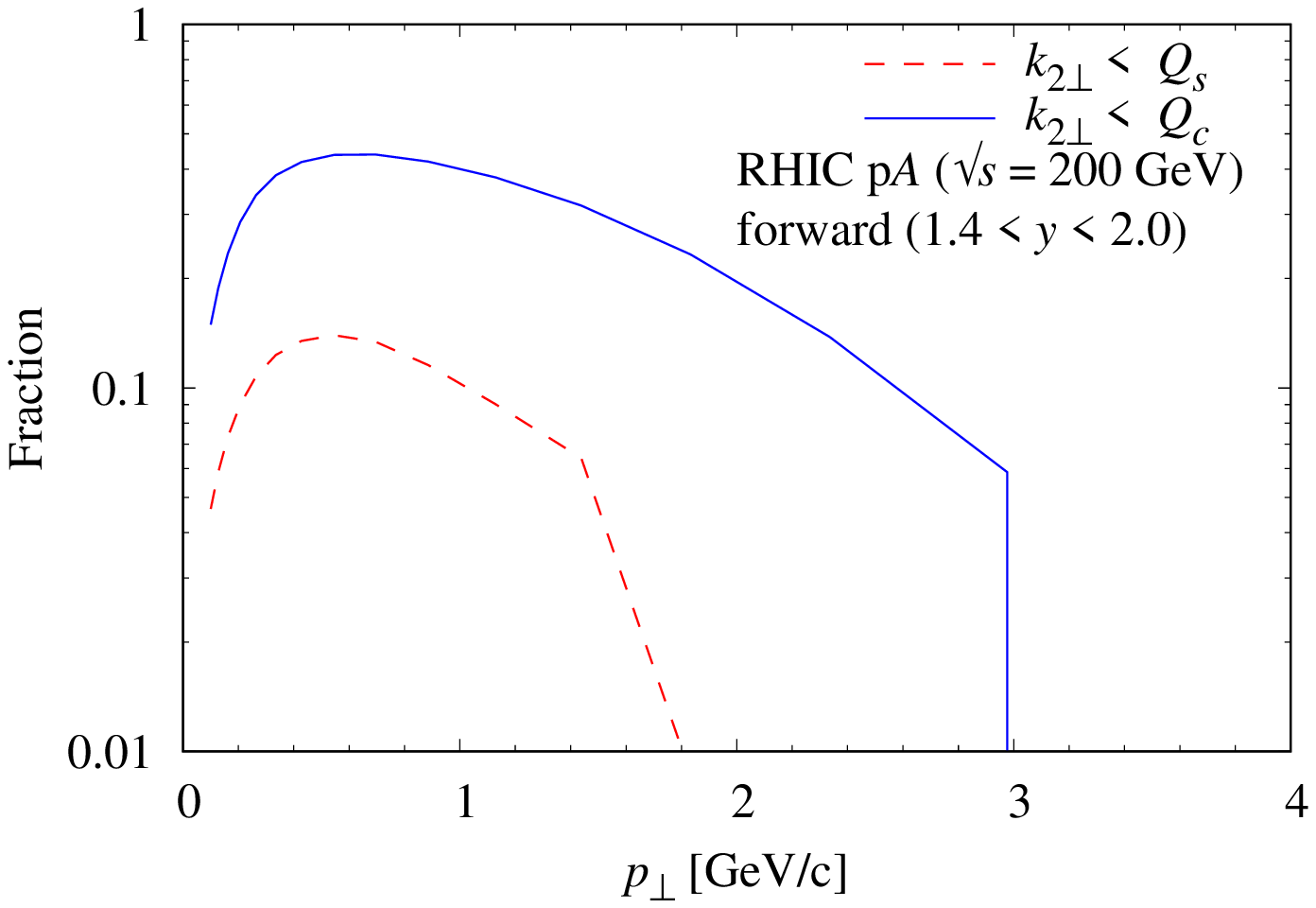} 
 \caption{Fractional contributions of saturation regime in muon yield from charmed-hadron decay in pp and p$A$ collisions at forward rapidity ($1.4<y<2.0$) at $\sqrt{s_{NN}}=200$ GeV: Red dashed and blue solid lines denote the contributions from the gluons inside the saturation region ($k_{2\perp}<Q_s$) and the extended scaling region ($k_{2\perp}<Q_s^2/\Lambda_\text{QCD}$), respectively. All the results are computed with Eq.~(\ref{eq:cross-section-LN-coll}) with $m_c=1.5$ GeV. }
\label{cgc-extended-RHIC}
\end{figure}

\begin{figure}
 \centering
 \includegraphics[height=7.9cm,angle=270]{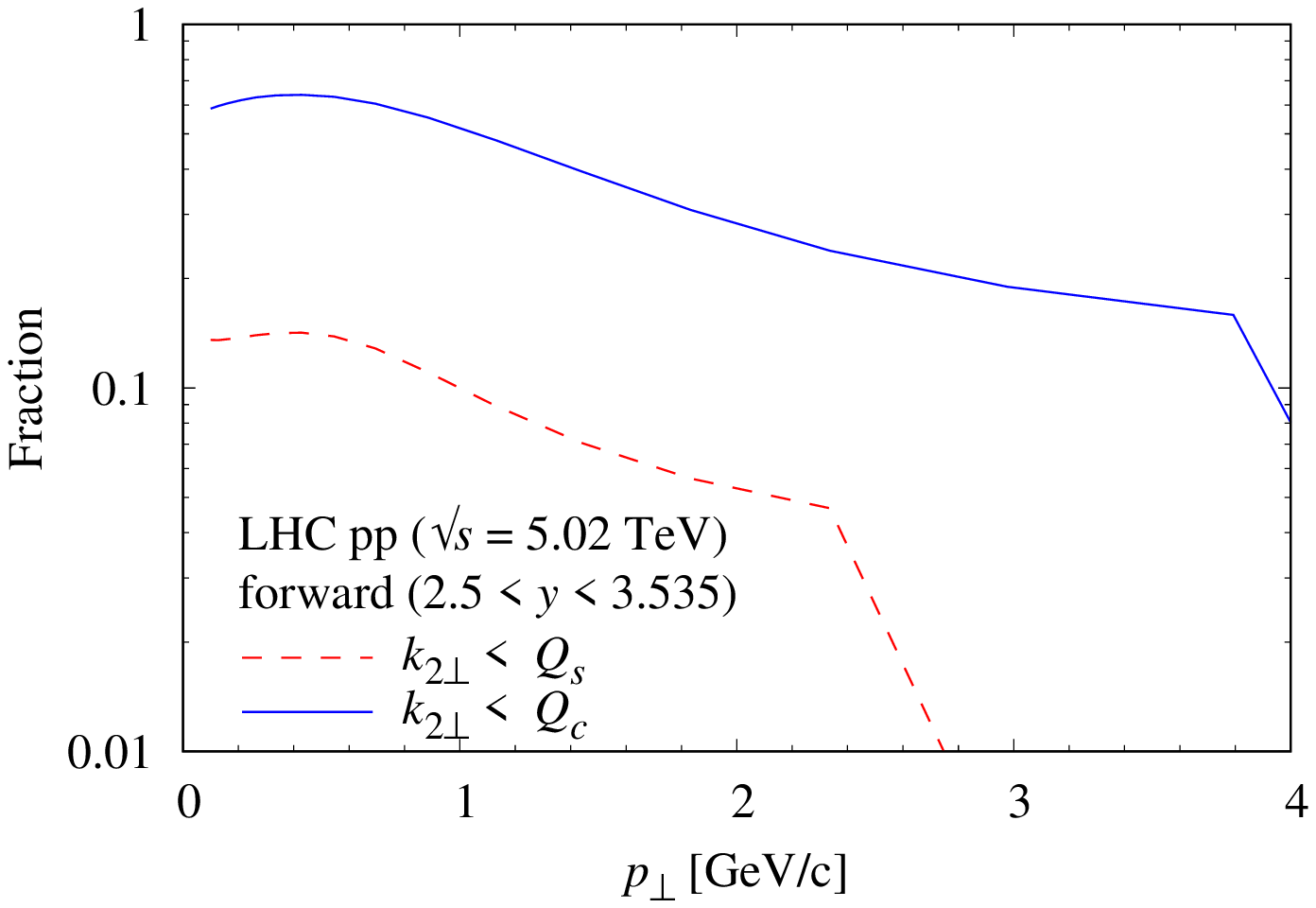} 
 \includegraphics[height=7.9cm,angle=270]{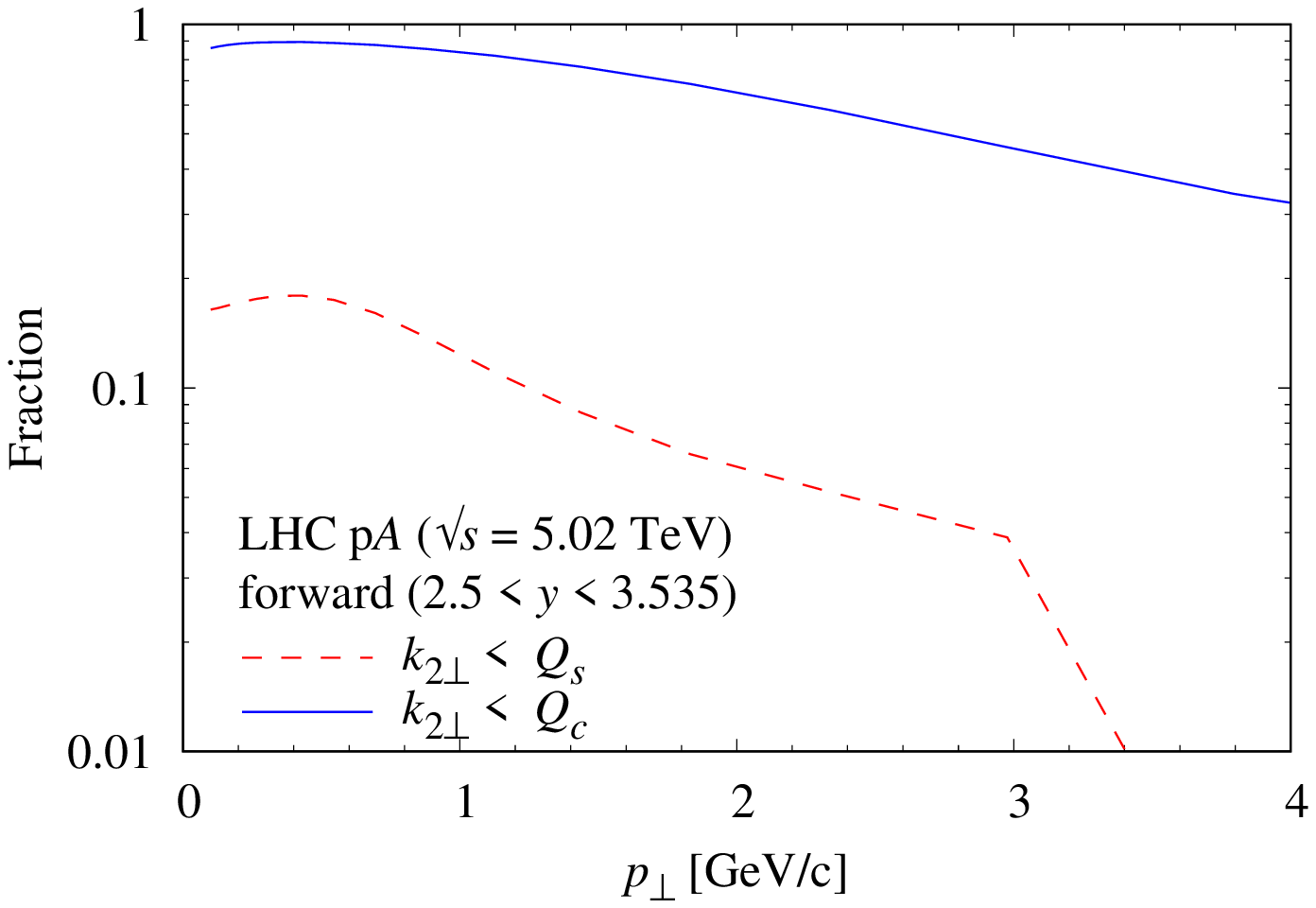} 
 \caption{Fractional contributions of saturation regime in muon yield from charmed-hadron decay in pp and p$A$ collisions at forward rapidity ($2.5<y<3.5$) at $\sqrt{s_{NN}}=5.02$ TeV. 
 Notations are the same as in FIG.~\ref{cgc-extended-RHIC}.}
 \label{cgc-extended-LHC}
\end{figure}

In order to quantify the saturation effects more precisely, we investigate fractions of the small-$x$ gluon contributions to the total spectrum in pp and p$A$ collisions.
Here, we introduce two cutoff scales in the integration variable $k_{2\perp}$ of the multipoint function $\phi$ in the integral:
One is the saturation scale $Q^2_s(x)$, which we determined by the condition
$S_Y(r_\perp=1/Q_s)=0.5$ with the solution of the rcBK equation, the other is the scale $Q^2_c(x)=Q_s^4(x)/\Lambda^2_\text{QCD}$ which
corresponds to the extended geometrical scaling line~\cite{Iancu:2002tr}. 
We will call the momentum regions of $k_{2\perp}<Q_s$ and $k_{2\perp}<Q_c$
the saturation region and the extended scaling region, respectively.
We use these representative scales in this work, and study the fractional contributions of these kinematical regions to the cross-section of the forward muon production, using
the hybrid formula~(\ref{eq:cross-section-LN-coll}) with MV$^\gamma$ parametrization.

FIG.~\ref{cgc-extended-RHIC} shows the fractions of the saturation and extended scaling regions to the differential cross-section of the muons from charmed-hadron decays at the RHIC energy.
In pp collisions the saturation region and the extended scaling region give only fractional contributions to the cross-section at low $p_\perp \lesssim 2$ GeV.
This is because the saturation scale of proton is small at RHIC. 
The muons at $p_\perp\gtrsim 3$ GeV 
are produced from the gluons outside the saturation regime.
In p$A$ collisions, the saturation scale of nucleus is enhanced. 
Then, the extended scaling region is pushed toward larger $p_\perp$ region 
although the saturation regime gives a fractional contribute to the total.

We show in FIG.~\ref{cgc-extended-LHC} the results at the LHC energy. 
One sees that both the saturation and extended scaling regions expand toward higher $p_\perp$ in pp and p$A$ collisions, compared to the results at RHIC. 
Remarkably, the cross-section in p$A$ collisions at $p_\perp\lesssim 1$ GeV is largely covered by the extended scaling region.

\subsection{Revising quarkonium and heavy flavor meson}

\begin{figure}
 \centering
 \includegraphics[height=7.9cm,angle=270]{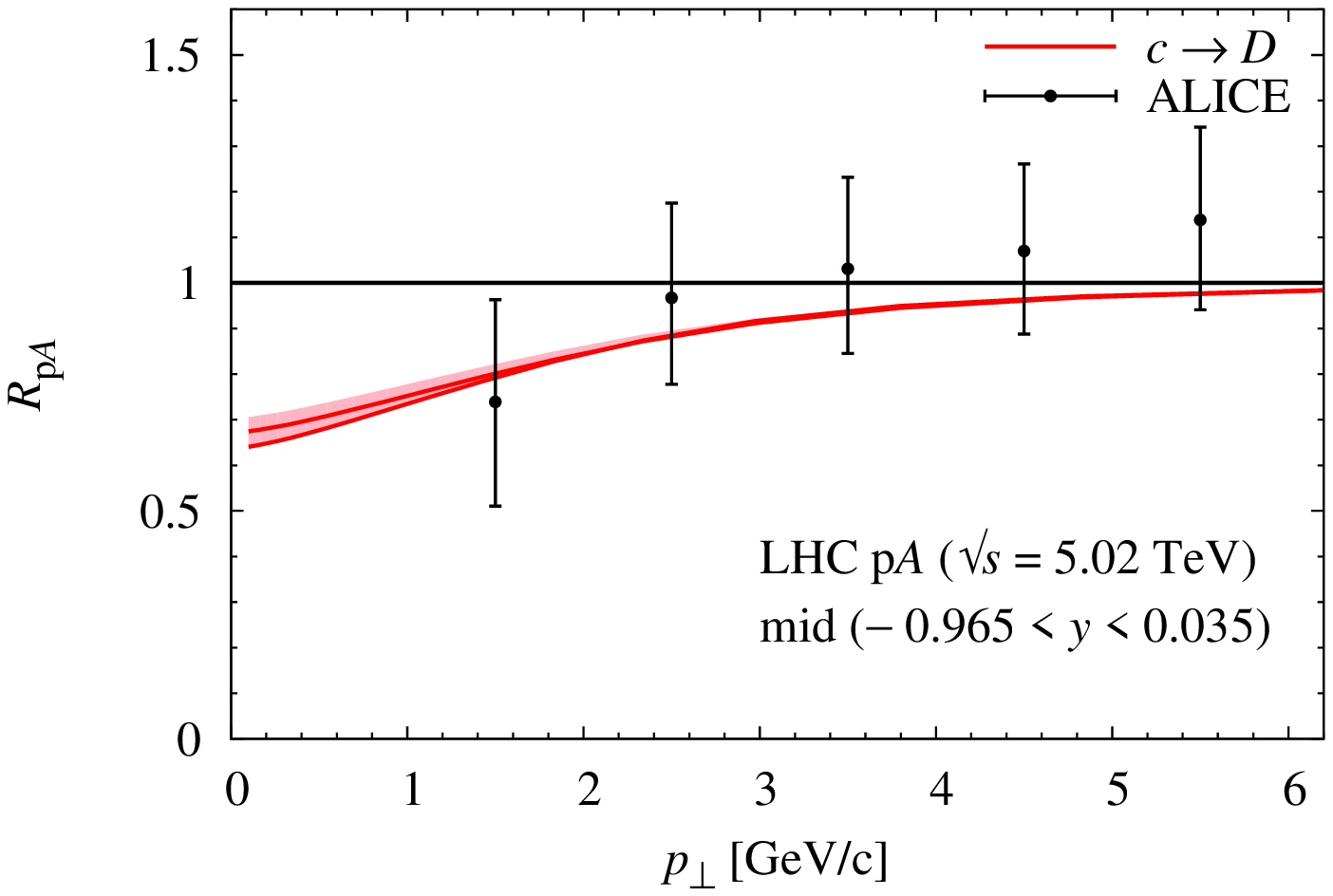} 
 \caption{Nuclear modification factor of $D$ meson production at $\sqrt {s_{NN}}=5.02$ TeV.
 Notations are the same as in FIG.~\ref{rpa-rhic}.
LHC data is taken from Ref.~\cite{Abelev:2014hha}.
 }
 \label{D-rpa-lhc}
\end{figure}

Finally, we re-evaluate $R_{{\rm p}A}$ of $J/\psi$ and $D$ meson productions, 
since we use $Q_{s0,A}^2=cA^{1/3}Q_{s0,p}^2$ with $c=0.5$ in the rcBK initial condition, which is a different value from what we used in our previous studies~\cite{Fujii:2013gxa,Fujii:2013yja}.
FIG.~\ref{D-rpa-lhc} shows the nuclear modification factor $R_{{\rm p}A}$ of $D$ meson as a function of $p_\perp$ at mid rapidity at the collision energy $\sqrt{s_{NN}}=5.02$ TeV. The result is in good agreement with the LHC data.
We find a slightly stronger suppression at low $p_\perp$ ($\lesssim 1$~GeV) compared to the results of the electrons. This is because the electrons at low $p_\perp$ come from the decays of the $D$ mesons with higher $p_\perp$.

\begin{figure}
 \centering
 \includegraphics[height=7.9cm,angle=270]{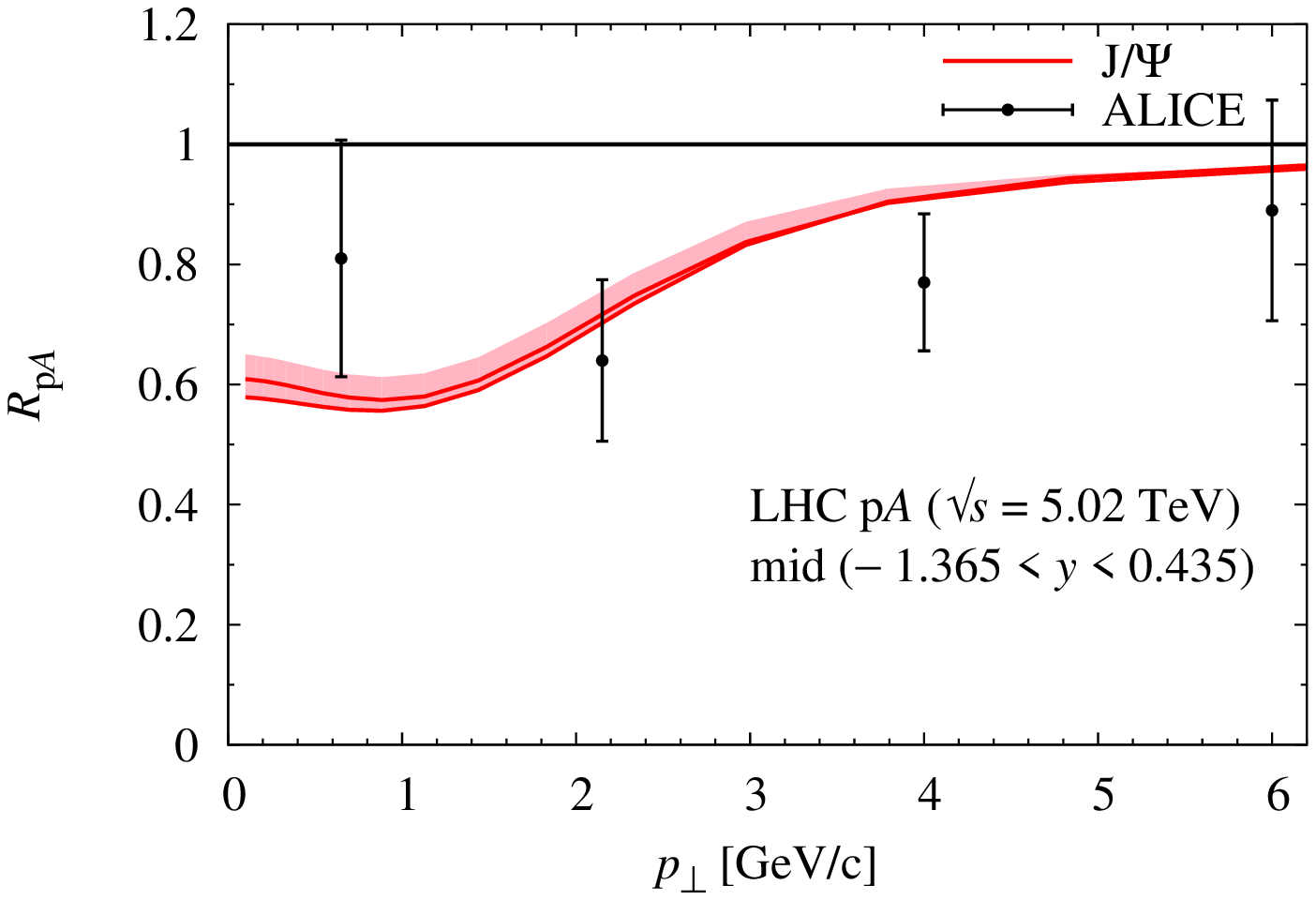} 
 \includegraphics[height=7.9cm,angle=270]{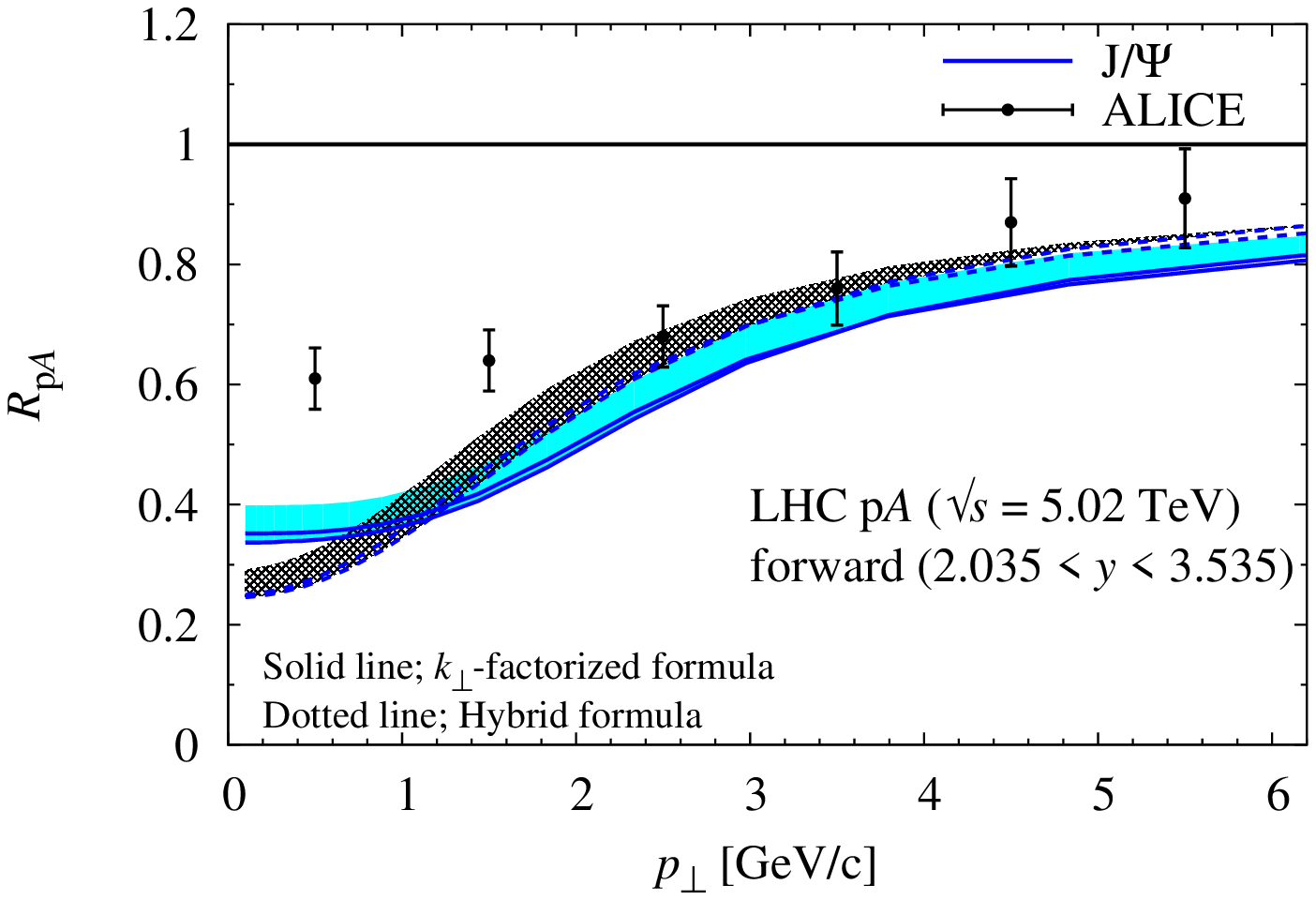} 
 \caption{Nuclear modification factor of $J/\psi$ production at $\sqrt {s_{NN}}=5.02$ TeV.
 Notations are the same as in FIG.~\ref{rpa-rhic}.
LHC data at mid and forward rapidities are taken from Ref.~\cite{Adam:2015iga}
 }
 \label{Jpsi-rpa-lhc}
\end{figure}

To the quarkonium production we adopt Color-Evaporation-Model (CEM) with the same parameters in Ref.~\cite{Fujii:2013gxa}. 
FIG.~\ref{Jpsi-rpa-lhc} shows $R_{{\rm p}A}$ of $J/\psi$ at mid and forward rapidities at the collision energy $\sqrt{s_{NN}}=5.02$ TeV.
Compared to our previous result of  $J/\psi$'s $R_{{\rm p}A}$ at forward rapidity~\cite{Fujii:2013gxa}, 
the new estimate comes closer to the experimental data as expected.
This may imply that the use of $Q_{s0,A}^2=cA^{1/3}Q_{s0,p}^2$ with $c=0.5$ in the initial condition of the rcBK equation for heavy nuclei is consistent for nuclear DIS and heavy quark production in
minimum bias events in p$A$ collisions.
We note that the value of $R_{{\rm p}A}$ for $J/\psi$ is quite similar to the recent calculation with taking into account the DIS constraint and the transverse profile of the nucleus by Duclou\'e et al.~\cite{Ducloue:2015gfa}.

Finally, by inspecting the whole results of the nuclear modification factors $R_{{\rm p}A}$ of $J/\psi$, $D$ meson and the electron, 
our calculation shows a hierarchy of $R_{{\rm p}A}^{l}>R_{{\rm p}A}^{D}>R_{{\rm p}A}^{J/\psi}$ at lower $p_\perp$.


\section{Summary}

In this paper, we have evaluated the lepton yields from charm-quark semileptonic decays
in pp and p$A$ collisions at the RHIC and LHC energies within the CGC framework.
Our result of the $p_\perp$ spectra of the heavy-quark decay leptons 
with the $k_\perp$ factorized formula (\ref{eq:cross-section-LN}) is harder than
the experimental data, but the hybrid formula (\ref{eq:cross-section-LN-coll}) 
gives rise to a consistent spectrum of the muons with the data in forward region
in pp collisions.
For the heavy nucleus, we adopted the initial saturation scale determined by 
$Q_{s0,A}^2=cA^{1/3}Q_{s0,{\rm p}}^2$ with the coefficient $c=0.5$, which is taken from the nuclear DIS analysis
\cite{Dusling:2009ni}.
We normalized the cross-sections in pA collisions so as to give 
the nuclear modification factor $R_{{\rm p}A}=1$ 
at large transverse momenta.
The $p_\perp$ dependence of $D$ meson (charm decay electron) production at mid rapidity in p$A$ collisions appears to be consistent with the data up to $p_\perp \sim 8$ GeV ($p_\perp \sim 4$ GeV).

We have shown that $R_{{\rm p}A}$ has weak $p_\perp$ dependence
in mid and forward rapidity regions at the RHIC energy, which can be understood as multiple scattering effects in the heavy nucleus.  
At the LHC energy,
$R_{{\rm p}A}$ is below unity at low  $p_\perp$ 
and recovering to unity at  higher $p_\perp$, 
due to the small-$x$ evolution of the gluon distribution in the target nucleus. 
Our results are in agreement with the data,
albeit there are large uncertainties.
The factor $R_{{\rm p}A}$ of the muons is predicted to be more suppressed 
at forward rapidity at the LHC energy.

We have examined the fractional contributions of the gluons in the nucleus 
in the saturation and extended scaling regions to the lepton yields.
At the RHIC energy, the contributions from these regions are very limited. On the other hand, at the LHC energy, the lepton yields at $p_\perp \lesssim 2$ GeV at forward regions in pA collisions are dominated by the contributions from the extended-scaling region ($k_\perp\lesssim Q_c$), and should reflect the saturation effects.

We have re-evaluated the nuclear modification factors
$R_{{\rm p}A}$ of J$/\psi$ and $D$ meson with the initial saturation scale
$Q_{s0,A}^2=cA^{1/3}Q_{s0,{\rm p}}^2$ with $c=0.5$ and obtained the results closer to the experimental data and consistent with \cite{Ducloue:2015gfa}.
From these results, we have found that
the nuclear modification factors of J$/\psi$, $D$ meson and the decay leptons show
suppression at forward rapidity at the LHC energy, reflecting the small-$x$ evolution 
of the gluon saturation effect.

In this work, we only considered
the minimum bias events in p$A$ collisions, but the study of the centrality dependence 
of the particle production will provide more information on the gluon saturation 
in the nucleus.
Possible  ways to treat the centrality dependence of collisions are the Glauber model with smooth nuclear thickness function~\cite{Ducloue:2015gfa} and the Monte Carlo implementation~\cite{Fujii:2011fh}. We leave this extension for future study.

The azimuthal angle correlation of the decay leptons may may provide further information of the saturation. In fact, we found in Ref.~\cite{Fujii:2013yja} that $D$-$\bar D$ correlation at away side ($\Delta\phi\sim\pi$) can probe the low $k_\perp$ gluon structure where the nonlinear effect is large. 
At the same time, one expects that the correlation of the two leptons from semileptonic decays in p$A$ collisions will be more smeared. In fact, back-to-back correlation of the pair at low $p_\perp$ can be sensitive not only to the saturation effect but also to the so-called Sudakov factor~\cite{Mueller:2013wwa} already in the quark production process.

Finally we note that the rcBK equation includes only a subset of NLO corrections while
the hard matrix elements at LO are adopted here in this phenomenological analysis
of pA collisions at the LHC energy.
Apparently a consistent NLO formulation, with NLO BK equation~\cite{Lappi:2015fma,Iancu:2015vea,Iancu:2015joa} and also the Sudakov resummation, is desired for more systematic analysis of the data.
For single quark production, the Sudakov effect will not be small. 
We leave it for future study.

\section*{Acknowledgements}
KW wishes to thank Komaba particle and nuclear theory group at the University of Tokyo for kind hospitality during his visit when this work was finalized. 
The authors would like to thank Bo-Wen Xiao for useful comments. 
The work of HF was partially supported by JSPS KAKENHI (\# 24540255).

\subsection*{Note added (Date : \today)}
In this update,  we have just fixed a typo in the published manuscript,
where a factor of $1/(2\pi)^2$ is missing in Eq.~(\ref{eq:cross-section-LN}).
This factor was correctly included in our numerical code,
and there is no change in the results.

\end{document}